\documentclass[jcp, aip, amsmath,amssymb, reprint]{revtex4-2}

\usepackage{natbib}
\usepackage{amssymb}
\usepackage{amsmath}
\usepackage[utf8]{inputenc}
\usepackage[T1]{fontenc}
\usepackage{mathptmx}
\usepackage[pdftex]{graphicx}
\usepackage{braket}
\usepackage{subfigure}
\usepackage{comment}
\usepackage[usenames,dvipsnames]{color}
\usepackage{hyperref}
\usepackage{dcolumn}
\usepackage{bm}     

\newcommand{\bld}{\boldsymbol}
\newcommand{\beq}{\begin{equation}}
\newcommand{\eeq}{\end{equation}}
\newcommand{\bea}{\begin{eqnarray}}
\newcommand{\eea}{\end{eqnarray}}
\newcommand{\bi}{\begin{itemize}}
\newcommand{\ei}{\end{itemize}}
\newcommand{\benum}{\begin{enumerate}}
\newcommand{\eenum}{\end{enumerate}}

\usepackage[normalem]{ulem}
\definecolor{darkgreen}{rgb}{0.0,0.66,0.0}
\definecolor{darkred}{rgb}{0.75,0.00,0.0}
\definecolor{darkblue}{rgb}{0.00,0.00,0.75}

\definecolor{URLCOL}{rgb}{0.1,0.2,0.7} 
\definecolor{LINKCOL}{rgb}{0.1,0.2,0.7} 
\definecolor{CITECOL}{rgb}{0.1,0.2,0.7} 
\definecolor{PREPRINTCOL}{rgb}{0.0,0.0,0.0} 

\begin{document}

\preprint{Redd and Cancio,  preprint 2020}

\title{ Asymptotic Analysis of the Pauli Potential for Atoms }
\author{Jeremy J. Redd}
\email{reddjer@uvu.edu}
\affiliation{Department of Physics, Utah Valley University, Orem, UT 84058}
\author{Antonio C. Cancio}
\affiliation{Department of Physics and Astronomy, Ball State University, Muncie, Indiana 47306}

\date{\today}

\begin{abstract} 
ABSTRACT:
Modeling the Pauli energy, the contribution to the kinetic energy caused by Pauli statistics, without using orbitals is the open problem of orbital-free density functional theory.
An important aspect of this problem is correctly reproducing the Pauli 
potential, the response of the Pauli kinetic energy to a change in density.
We analyze the behavior of the Pauli potential of non-relativistic neutral atoms under 
Lieb-Simon scaling -- the process of taking nuclear charge and particle number 
to infinity, in which the kinetic energy tends to the Thomas-Fermi limit.
We do this by mathematical analysis of the near-nuclear region and by calculating the exact 
orbital-dependent Pauli potential using the approach of Ouyang and Levy for closed-shell atoms 
out to element Z=976.  
In rough analogy to Lieb and Simon's own findings for the charge density, we find that the
potential does not converge smoothly to the Thomas-Fermi limit on a point-by-point basis
but separates into several distinct regions of behavior.
Near the nucleus, the potential approaches a constant given by the difference
in energy between the lowest and highest occupied eigenvalues. 
We discover a transition region in the 
outer core where the potential deviates unexpectedly and predictably from both
the Thomas-Fermi potential and the gradient expansion correction to it.
These results may provide insight into semi-classical description of Pauli statistics, and new 
constraints to aid the improvement of orbital-free DFT functionals.
\end{abstract}


\maketitle

\section{Introduction}\label{Introduction}

	The most generally accurate and widely used method for
predicting electronic structure is the Kohn-Sham (KS) approach to
density functional theory (DFT).~\cite{KS}
By introducing auxiliary orbitals into the definition of particle density,  
the KS functional allows for an accurate representation of the
energy of the exact many-body Hamiltonian by the energy of a simpler
noninteracting system.~\cite{Martin, IRG}
This greatly simpliifies the mathematics and speeds up computations
as compared to many-body or Hartee-Fock calculations.~\cite{karasiev}  
However, the use of orbitals still comes with increasing computational cost as
the number of particles in the system is up-scaled.
This means that for systems that require the calculation of many orbitals
such as mesoscale systems where quantum properties may be 
important\cite{akimov2015large} 
and warm dense matter,~\cite{graziani2014frontiers, rosnerbasic, karasievWDM}
in which many states become thermally activated,
the computational cost of the KS method becomes prohibitive.

 
 	The Hohenburg-Kohn theorem, however, states that the ground state of 
any Hamiltonian system can be uniquely characterized by the particle density
alone.~\cite{HK}  This means that exact Hamiltonian solutions can be expressed as functionals of exclusively the density, eliminating the need 
for orbitals.\cite{karasiev,TranWes}
Crucially, this theorem applies to any piece of the  
energy, so not only is the true interacting Kinetic energy a
functional of the density but also the KS kinetic energy:
as conventionally defined, this is a functional of the KS orbitals,
nevertheless a more general orbital-free expression should exist.  


Recent years have seen a growing number of approaches to constructing orbital-free DFT (OFDFT) 
approximations to the KS method that allow for improved computational scaling for systems with 
large numbers of orbitals.\cite{TranWes, witt2018orbital} The orbital-free philosophy has also 
been introduced successfully into 
conventional KS DFT, in the form of ``de-orbitalizing'' exchange-correlation 
functionals that depend explicitly upon the KS kinetic energy density (KED), 
replacing it with an equivalent expression in terms of the density and 
its derivatives.~\cite{trickey_deorbitalization}

However the challenge of developing a robust OFDFT model with reasonable 
predictive accuracy for a variety of systems is severe.
In conventional Kohn-Sham DFT one needs to approximate 
the exchange-correlation energy describing the difference 
in energy between interacting and noninteracting systems for the 
same external potential, normally a small correction.
OFDFT must approximate the kinetic energy, which is of the order
of the energy itself   
and must therefore be modeled to high accuracy.

The most basic OF theory, Thomas-Fermi (TF) 
theory,~\cite{thomas1927calculation, F27}
uses the KE of the homogeneous electron gas applied to the local density, 
in analogy to the LDA of the KS method.	 
But unlike the LDA, which produces at least qualitatively good
structural predictions, TF theory does not permit chemical 
binding at all.~\cite{teller,cartermol,finzel2018chemical}
The simplest functional beyond Thomas-Fermi, 
the gradient expansion (GE), does very well for atoms, but still not
so well for molecular binding. 
Many attempts have been made to build on this foundation to develop 
semilocal or ``single-point'' functionals using the local density 
and its gradient as ingredients,~\cite{TranWesolowski, LacksGordon94, 
Thakkar92, APBE, VT84F, BorgooMol, LKT}
sometimes adding the Laplacian of the 
density,~\cite{PC,LCPB,cancioredd,CFS2018,CFS20194thorder} and 
the electronic Hartree potential.~\cite{constantin2017modified} 
These more complex models generally share the problems of their predecessors,
but can be competitive~\cite{CFS2018} with more expensive empirical 
nonlocal functionals for some solids. 

Two-point nonlocal functionals have been somewhat more 
successful.~\cite{WangCarterNew,huangcarter,pavanello}
These are based on the 
Lindhard formula for linear response of the homogeneous
electron gas.
However, the Lindhard function is not an appropriate reference
point for finite systems
and systems with surfaces.~\cite{ConstLindhard}  
At least to date, 
such functionals require system-dependent empirical parameters to succeed.

The challenge of OFDFT is modeling the kinetic energy due to the Pauli 
exclusion principle -- the orbital dependence in the KS functional is a 
consequence of Pauli statistics.  
One considers the total kinetic energy as a sum of this Pauli KE contribution
and the von Weizs\"{a}cker KE -- the KE 
of a fictitious Bose system with the same density as the real system.  
Minimizing the OFDFT energy then generates
an Euler equation for the density for this fictitious Bose system 
where the contribution of Pauli statistics appears as an effective
Pauli potential, that forces this fictitious system to have the 
same density as the true fermionic one.
This Pauli potential thus is analogous 
to the Kohn-Sham potential for the conventional Kohn-Sham method.

The Pauli potential thus plays an important role
in guaranteeing the stability and accuracy of structural calculations;
nonetheless, like the Kohn-Sham potential, it gets much attention in
developing functionals than the Pauli energy. 
A notable exception to this tendency is the use of the non-negativity of the potential 
as a constraint -- a significant feature
of at least one family of functionals.~\cite{VT84F, LKT} 
Nonetheless, a quite pleasing property of the 
exact Pauli potential is that it can be easily constructed in terms of KS orbitals in a simple fashion.\cite{levy1988exact}  Essentially one can use the 
orbital definition of density to solve the KS problem and equivalent Euler problem simultaneously.  This allows one to compare the results 
of model Pauli potentials to the exact potential for any system of interest.
Exact Pauli potentials have been constructed in this way for example
atoms,~\cite{levy1988exact, gritsenko1994, van1995step, baerends1997quantum, KraislerAxel20, 
FinzelMolecules20}
Approximate Pauli potentials play a key role in a recently developed OF 
method~\cite{finzel2018chemical, Finzel19, FinzelMolecules20}. 
These rely on the orbitals of isolated atoms and an 
orbital-free description of the bond, thus constituting a hybrid approach to deorbitalizing the 
KS problem.


One way to generate useful constraints for functional development -- whether
on the total energy or the potential -- is to consider the behavior of the 
functional under scaling of the system.
A particularly fruitful example is 
Lieb-Simon scaling, the best
known example of which is the scaling of the KE of neutral atoms as nuclear charge 
tends to infinity.~\cite{liebsimon, BCGP16,LCPB}
This should not be confused with Levy-Perdew scaling which more closely 
resembles the scaling of nuclear charge to infinity with constant 
particle number.\cite{levyperdew}  
The lower bound of this scaling behavior is the von Weizs\"acker (VW) solution for 
hydrogen and helium, trivially convertible to orbital-free form because 
it involves only one occupied orbital.  
The upper bound is less simple but more powerful.  The leading order in $Z$ of the 
total and kinetic energy as $Z\to\infty$ is given by TF theory~\cite{liebsimon}.  
The gradient expansion approximaton (GEA) contributes corrections of smaller
order in $Z$ to the large $Z$ limit of the energy.~\cite{BCGP16,LCPB} 
The limiting behavior of total energies is reflected in the kinetic energy 
density, which is locally approximated by a variant of the gradient 
expansion in the core region of the atom.~\cite{cancioredd}

The physical property that has not been explored carefully in the Lieb-Simon limit
is the Pauli potential.  
Even though the total Pauli kinetic energy should be well described by 
TF theory in this limit, the same does not 
necessarily hold point for point for the potential. And it is unknown to what
extent functionals that are  successful in describing total energies
work for the potential.
%
%
%
%

In this paper, we analyze the behavior of the exact KS Pauli potential 
for nonrelativistic neutral atoms as a function of $Z$ up to $Z=976$, large
enough to extract limiting behavior 
and exact constraints that may be of aid to the development of KE functionals. 
We find an exact constraint in the near-nucleus limit 
and an unexpected deviation from the Thomas Fermi limit for a 
the outer shells of the large-$Z$ atom.   
	
	The paper is organized as follows: 
	 Section~\ref{Theory} discusses the theoretical background of Pauli potentials, both in the context of KS and of OFDFT approximations.  
	 Section~\ref{Methods} describes the methods and algorithms used for calculations and validation of the results.
	 Section~\ref{Results} details the visual results of extending the exact Pauli response functionals and Pauli potentials to large-$Z$, as compared
to approximations.
	 Section~\ref{Conclusions} discusses the 
ramifications of our findings and possible future work.  
	 
	\section{Theory} \label{Theory}
In Kohn-Sham theory, the total energy of an electronic system as
a functional of the density $\rho$ is given by
	\beq
		\label{eq:Eks}
		E[\rho]=T_{KS}[\rho]+U[\rho]+\int v_{ext}(\bld{r})\rho(\bld{r}) d^3r+E_{xc}[\rho],
	\eeq
	where $T_{KS}[\rho]$ is the noninteracting contribution to the KE, $U[\rho]$ is the static electron-electron interaction, $v_{ext}$ is an external potential, and $E_{xc}$ is the energy of exchange and correlation effects.  The last term contains the difference in energy 
between the true interacting system and the fictitious noninteracting one.
	
	The KS density is given by
	\beq
		\label{eq:rhoks}
		\rho(\bld{r})=\sum_{i}f_{i}|\phi_{i}(\bld{r})|^2,
	\eeq
where $\phi_i$ are the auxiliary single-particle orbital that describe the noninteracting system and
$f_i$ is the occupation number.  
The KS kinetic energy is then given by
\beq
\label{eq:tauks}
T_{KS}=\int\tau_{KS}d^3r
=\int\frac{1}{2}\sum_{i}f_{i}
|\nabla \phi_{i}(\bld{r})|^2 d^3r,
\eeq
where $\tau_{KS}$ is a positive-definite kinetic energy density.

The density is determined 
by the functional minimization of the energy with respect 
to each orbital, with the constraint 
of preserving orbital normalization.  This generates the effective Kohn-Sham equation for each orbital:
	\beq
		\label{eq:eks}
                \left[\frac{1}{2}\nabla^2+v_{KS}(\bld{r})\right]\phi_{i}(\bld{r}) = \epsilon_{i}\phi_{i}(\bld{r})
	\eeq
where $\epsilon_i$ is an auxiliary eigen value.
The KS potential is determined from the functional derivative of the energy:
	\beq
	\label{eq:vsham}
	v_{KS}=\frac{\delta}{\delta\rho}\left(U[\rho]+E_{xc}[\rho]\right)+v_{ext}.
	\eeq

In order to generate an orbital-free version of the Kohn-Sham functional, 
we define the Pauli KE
as the difference between KS and vW kinetic energies.
\beq
\label{eq:Tp}	
T_{p}=T_{KS}-T_{vW},
\eeq
and define a Pauli KE density, the integral over which yields the Pauli KE, similarly:
\beq
\label{eq:taup}	
\tau_{p}=\tau_{KS}-\tau_{vW}.
\eeq
As discussed in the introduction, the vW kinetic energy, in the spirit of 
the KS idea, is the kinetic energy of a fictitious
Bose system that has the same energy and density as the true, fermionic system.
In this case, all particles occupy the ground state, $\rho = N|\psi_0|^2$, 
so that the associated KE density is
\beq
\label{eq:tauvw}
\tau_{vW}=-\frac{1}{2}\left| \nabla\sqrt{\rho(r)}\right|^2=
\frac{1}{8}\frac{|\nabla \rho(\bld{r})|^2}{\rho(\bld{r})}.
\eeq
This is strictly correct for the true system only if $N\leq 2$.
	The Pauli KE then measures the additional kinetic
energy due to Fermi statistics.  This has to be approximated somehow by a functional
of the density, in a way similar to how the XC energy incorporating 
electron interactions is approximated in KS theory. 

One can now generate an orbital-free Euler expression of the KS problem. 
It calculates the non-interacting Bose KE explicitly and considers effects of the Pauli contribution to the KE 
to come from a positive definite Pauli potential $v_p$.  
Minimizing $E - \mu \int \rho(\mathbf{r}) d^3r$, one finds
\beq
\label{eq:tsqrtrho}
\left[-\frac{1}{2}\nabla^2+v_\mathrm{eff}(\mathbf{r})\right]\sqrt{\rho(\mathbf{r})}=\mu\sqrt{\rho(\mathbf{r})}.
\eeq
Like the KS procedure, this generates an effective potential $v_\mathrm{eff}$, and solves for the
density and a single eigenvalue $\mu$, the chemical potential. 
The effective potential is given by $v_\mathrm{eff}=v_{KS}+v_p$, with 
the addition to the Kohn-Sham potential -- the Pauli potential $v_p$ --
given by\cite{levy1988exact} 
\beq
\label{eq:vp}
v_{p}(\mathbf{r})=\frac{\delta T_{p}[\rho]}{\delta\rho(\bld{r})}.
\eeq
It can be interpreted as the potential needed to make the density of the 
fictitious Bose system calculated with Eq.~[\ref{eq:tsqrtrho}]
equal the density of the fermionic Kohn-Sham system.

The Pauli potential may be determined exactly in terms of the KS orbitals as~\cite{levy1988exact}
\beq
\label{eq:vpauli}
v_p(\mathbf{r})=\frac{\tau_p(\mathbf{r})}{\rho(\bld{r})}+v_{r}(\mathbf{r})
\eeq
where $v_{r}$ is the response of the effective potential, and consequently the KE, to an arbitrary change in density.  The exact response potential is given by \cite{levy1988exact}
\beq
\label{eq:vrespexact}
v_r(\mathbf{r})=\frac{2}{\rho(\bld{r})} \sum_{j=1}^{M}(\epsilon_{M}-\epsilon_{j})\phi_{j}^{*}(\bld{r})\phi_{j}(\bld{r}).
\eeq  
Eq.~\ref{eq:vpauli} and~\ref{eq:vrespexact} can be derived by simultaneously 
solving Eq.~\ref{eq:eks} and Eq.~\ref{eq:tsqrtrho} using the same density Eq.~\ref{eq:rhoks} 
where the occupation $f_i =2$. 

The primary tool for our study of atomic Pauli potentials is Lieb and Simon's
$\zeta$ scaling of the kinetic energy of 
neutral atoms~\cite{liebsimon, lieb1977thomas, BCGP16, LCPB, cancioredd}.
~The Lieb-Simon theorem scales the potential and particle number of a system simultaneously:
	\bea
		\label{eq:densscale}
		N_\zeta&=&\zeta N_1, \\
		\label{eq:potscale}
		v_\zeta(r)&=&-\frac{\zeta}{r}.
	\eea
This yields the neutral atoms for integer values $Z$ of the continuous variable 
$\zeta$ with the choice $N_1=1$.
Particle distance is then scaled in units of the  Thomas-Fermi atomic radius $\sim Z^{-1/3}a_0$
so that formally, the potential scales as 
	\beq
		\label{eq:volscale}
		v_\zeta=\zeta^{4/3} v_1(\zeta^{1/3}r).
	\eeq

The key result for this paper is that in the limit $\zeta \to \infty$, 
(for atoms, $Z\to \infty$) the total energy and 
thus also kinetic energy in the Thomas-Fermi approximation becomes 
relatively exact:
\beq
   \lim_{\zeta\to\infty} \frac{T_{KS} - T_{TF}}{T_{KS}} \to 0.
		\label{eq:LiebSimonlimit}
\eeq
Secondly, in the case of atoms, 
the TF energy and the leading corrections in the $Z\to \infty$ limit 
are exactly known and form an expansion in powers of $Z^{1/3}$:
	\beq
		\label{eq:TofZexpansion}
		T_{KS}=c_0 Z^{7/3}+c_1 Z^2+c_2 Z^{5/3}+...
	\eeq
Here the leading order $c_0=0.768745$ is predicted by 
TF theory,~\cite{thomas1927calculation}
$c_1=-1/2$,~\cite{scott1952, schwinger1980thomas} and $c_2=0.269900$.~\cite{schwinger1981thomas}
Any candidate for an orbital free KE functional ought to satisfy this 
scaling behavior,
but this not a trivial task.~\cite{LCPB}  The second correction $c_2$
is generated by the standard gradient expansion.  However,
the Scott correction, which scales as $Z^2$, is a larger effect and although it
may be modeled with a gradient expansion, it explicitly deals with the 
KE near the Coulomb singularity, where the GE is 
not legitimate.  Not surprisingly, very few GGA's or metaGGA's get this 
limit correctly.~\cite{PC,LCPB}

Given the importance of the gradient expansion model for the large $Z$ 
expansion, we will compare our results to functionals of this form; 
keeping in mind its limitations, we explore a number of variations on the
theme.
The leading order term of the large-$Z$ expansion is, as per Eq.~(\ref{eq:LiebSimonlimit}), given by the 
the Thomas-Fermi KED~\cite{spruch1991pedagogic} -- the KED in 
the limit of a homogeneous electron gas, applied to the local density $\rho(\bld{r})$:
		\beq
			\label{eq:tautf}
			\tau_{TF}(\bld{r})=\frac{3}{10}k_F(\bld{r})^2\rho(\bld{r})=\frac{3}{10}(3\pi^2)^{2/3}\rho(\bld{r})^{5/3}.
		\eeq
(It should be noted that this and subsequent model equations are defined 
for the Kohn-Sham and not the Pauli KED).
The subsequent orders depend on the gradient expansion of the kinetic energy for
the slowly varying electron gas.
The gradient expansion may be formally derived as an expansion in orders
of $\hbar$, good for large values of the local fermi energy -- in effect,
large numbers of occupied states.
To second order it is given by~\cite{kirzhnitsND}
	\beq
		\label{eq:taugea2}
		\tau_{GEA}=\left[1+\frac{5}{27}p+\frac{20}{9}q\right]\tau_{TF},
	\eeq
	where
	\beq
		\label{eq:p}
		p=\frac{|\nabla \rho|^2}{4{k_F}^2{\rho}^2},
	\eeq
	and
	\beq
		\label{eq:q}
		q=\frac{\nabla^2 \rho}{4{k_F}^2\rho}.
	\eeq
The fourth order~\cite{hodges73} correction improves on this 
model for atoms~\cite{LCPB,JandG} but will not be considered in this paper.

Although this ``canonical" GEA yields a reasonable description of the
large $Z$ expansion, it is not perfect.   
It neither appears to be the best candidate for describing the total 
KE of atoms~\cite{LCPB} nor the 
local KED.~\cite{lindmaa14,cancioredd}
In fact, a modification of the GEA ($\mathrm{Loc}$) can be 
determined by fitting the local KED of the core shells of 
large-$Z$ atoms~\cite{cancioredd}, which yields 
      \beq
      \label{eq:gealoc}
          \tau_{Loc}=\left(1-0.275p+2.895q\right)\tau_{TF}.
      \eeq
Although this does not yield good total energies, it is of obvious interest to model 
the \textit{potential}
which is also a local quantity and is related to the KED by Eq.~\ref{eq:vpauli}.
It is noteworthy that the gradient term of the $\mathrm{Loc}$ GE has
a sign opposite to that of the canonical GE, and thus a net correction 
to the TF energy which is negative, rather than positive.
Ref.~\onlinecite{cancioredd} also introduces a 
$Z$-dependent near-nuclear correction to this
that does yield good total energies, labelled $\mathrm{NNloc}$ in the results. 

A similar local model is that of Lindmaa, Armiento and Mattsson,~\cite{lindmaa14} given
by:
	\beq
            \label{eq:tauairy}
            \tau_{Airy}=\left[1-\frac{5}{27}p+\frac{30}{9}q\right]\tau_{TF},
	\eeq
This gradient expansion is derived from the analysis of 
the ``edge electron gas" or Airy gas~\cite{KohnMattsson} that is 
constructed by taking a linear potential with hard wall boundary, in the 
limit that the hard wall is moved to infinity.  Thus it is meant to be valid
for surfaces, and not necessarily as global functional.  
It has been shown to be a good approximation for the KED of model systems
including jellium droplets and the Bohr atom. 
As an atom
is necessarily a system with a surface region, it is of interest to see 
how it fares here.

A final variant of the GE is introduced by Tsirelson et al. (Ref.~\onlinecite{Ts}) which uses an estimate of the chemical
	potential to modify the large-$r$ limit. The most relevant portion of their model 
is the response function which is fit to the following form:
\beq
\label{eq:TSprefit}
v_{r,Ts} = \frac{(3\pi^2)^{2/3}}{5}\rho^{2/3}+a\frac{|\nabla\rho|^2}{\rho^2}+b\frac{\nabla^2\rho}{\rho},
\eeq
where $a=0.05$ and $b=0.14$.

We can now consider the functional derivative of the gradient expansion KE, 
to generate gradient expansion formulae for $v_{p}$ and $v_r$.
The kinetic energy using second-order differentials of the density may be written as
	\beq
		\label{eq:tksapprox}
		T^{approx}_{KS}=\int\tau^{approx}_{KS}\left[\rho(\bld{r}),\nabla \rho(\bld{r}), 
                                                            \nabla^2\rho(\bld{r})\right]d^3r.
	\eeq
The functional derivative of this form is
	\beq
		\label{eq:vksapprox}
                \frac{\delta T^{approx}_{KS}}{\delta \rho} = 
		\frac{\partial\tau_{KS}^{approx}}{\partial\rho}-\nabla\cdot\frac{\partial\tau_{KS}^{approx}}{\partial\nabla\rho}+\nabla^2\frac{\partial\tau_{KS}^{approx}}{\partial\nabla^2\rho}.
	\eeq
Now, consider an arbitrary second order GE of the form
	\beq
		\label{eq:genge}
		T^{approx}_{KS} = \int \left(1+\eta_Q q+\eta_P p\right )\tau_{TF} \,d^3r
	\eeq
To get the Pauli KE, we
subtract the von Weizsacker kinetic energy $\int (5p/3) \tau_{TF} \,d^3r$
from Eq.~(\ref{eq:genge}). 
Then applying the functional derivative [Eq.~(\ref{eq:vksapprox})]
yields the GE approximation of the Pauli potential:
\beq
\label{eq:vpauliGE}
v_{p}^{GE}=\left[\frac{5}{3} + (\eta_P-5/3) p - 2(\eta_P-5/3) q\right]\frac{\tau_{TF}}{\rho}.
\eeq
which is independent of $\eta_Q$ because the Laplacian 
term $\sim q$ does not contribute to the KE or its functional derivative.  
The response function follows trivially from Eq.~(\ref{eq:vpauli}):
	\beq
             \label{eq:vrespGE}
             v_{r}^{GE}=\left[\frac{2}{3} - (\eta_Q+ 2(\eta_P-5/3)) 
                    q\right]\frac{\tau_{TF}}{\rho}.
	\eeq

We finish by considering the regions of general behavior in atomic electron densities for large-$Z$ atoms proposed by the analysis of 
Lieb and Simon~\cite{liebsimon} and augmented by Heilman and Lieb.~\cite{heilmannlieb}
Moving outward from the center, there is first a region near the nucleus where TF behavior breaks down, consisting
of electrons whose behavior can be described described by the Bohr atom of non-interacting electrons.~\cite{heilmannlieb}
There is an inner core region, the density of which should behave as a slowly varying  electron gas obeying TF theory [Eq.~(\ref{eq:tautf})].  
The characteristic length of this core scales as $Z^{-1/3}$ and 
the density scales as $Z^2$.  
There is a ``mantle of the core" also with length scale $Z^{-1/3}$, and in which the density decays as $1/r^6$.
In the infinite-$Z$ atom the ratio of electrons outside core and mantle 
to those inside drops to zero. 
Then, there is a ``complicated transition region''~\cite{liebsimon} and a valence 
region of outer shells with a length scale presumably of order 1.  
Finally there is an evanescent region where the electron density decays exponentially.  
One check of how well we have approached the $Z\to\infty$ limit may be how many of these regions we can actually detect in our data.
	
	\section{Methods} \label{Methods}
In order to calculate Kohn-Sham orbitals and eigenvalues needed for the 
calculation of exact KEDs and response potentials we use
the atomic code FHI98PP\cite{FHI98PP} 
in its all-electron, non-relativistic mode. 
FHI98PP computes wave functions on a logarithmic grid, with spacing 
between successive points increasing by a geometric factor $\gamma$.  
We use the default  $\gamma=0.0247$ which yields inappreciably different 
results from 0.0123.
For simplicity, the exchange-correlation functional used was the PW91
LSDA.\cite{numGGA}  
The disagreement between LDA and exact Kohn-Sham calculations is known
to disappear in the large-$Z$ limit;~\cite{BCGP16} in practice, 
our kinetic energies agree with exact OEP calculations 
within 0.67\% for Ne and 0.055\% for Rn.  See Supplemental Material
for further details.
For differentiation of functions we employ Lagrange interpolating polynomials, similar to Gauss quadrature, with polynomials up to twelfth-order, while
for integration we use the composite Simpson's method.  Details are given
in Ref.~\onlinecite{thesis}. 

The construction of extremely large atoms should also be discussed.  
For the nonrelativistic case, one naively extends the Aufbau principle out to infinity.\cite{pyykko}  
For atoms with highly degenerate valence energy shells, the lanthanum and actinium series for instance, this is likely a poor assumption, because
completing such a shell might take preference over filling a lower energy
shell with low degeneracy. 
However for atoms in the eight principal columns of the periodic table, all 
highly degenerate shells are already completely filled and thus do not 
influence the filling order.

We have extended FHI98PP using the Aufbau principle out to element number 976,
with a 16p valence shell.  
The validity of this extension has been tested by comparing the total energy of Aufbau-constructed shells versus several other shell configurations for 
elements 976 (filled 16p), 970 (filled 15d), and 816 (filled 16s).
For all cases tested, the Aufbau construction proves to be the nonrelativistic 
ground state for these atoms.  
To check the quality of our numerical solutions, 
completely independent calculations were done with a second 
atomic DFT code, OPMKS,~\cite{opmks} for atoms with $Z<400$.  
The results are indistinguishable with those of FHI98PP within machine error.
A table of highest occupied atomic orbital (HOAO) eigenvalues and 
kinetic energies of large $Z$ atoms
from both methods is given in the Supplemental Material.

\section{Results} \label{Results}
	\subsection{Verifying densities}
\begin{figure}[!htbp]	\includegraphics[width=1.0\linewidth,height=0.41\textheight,keepaspectratio]{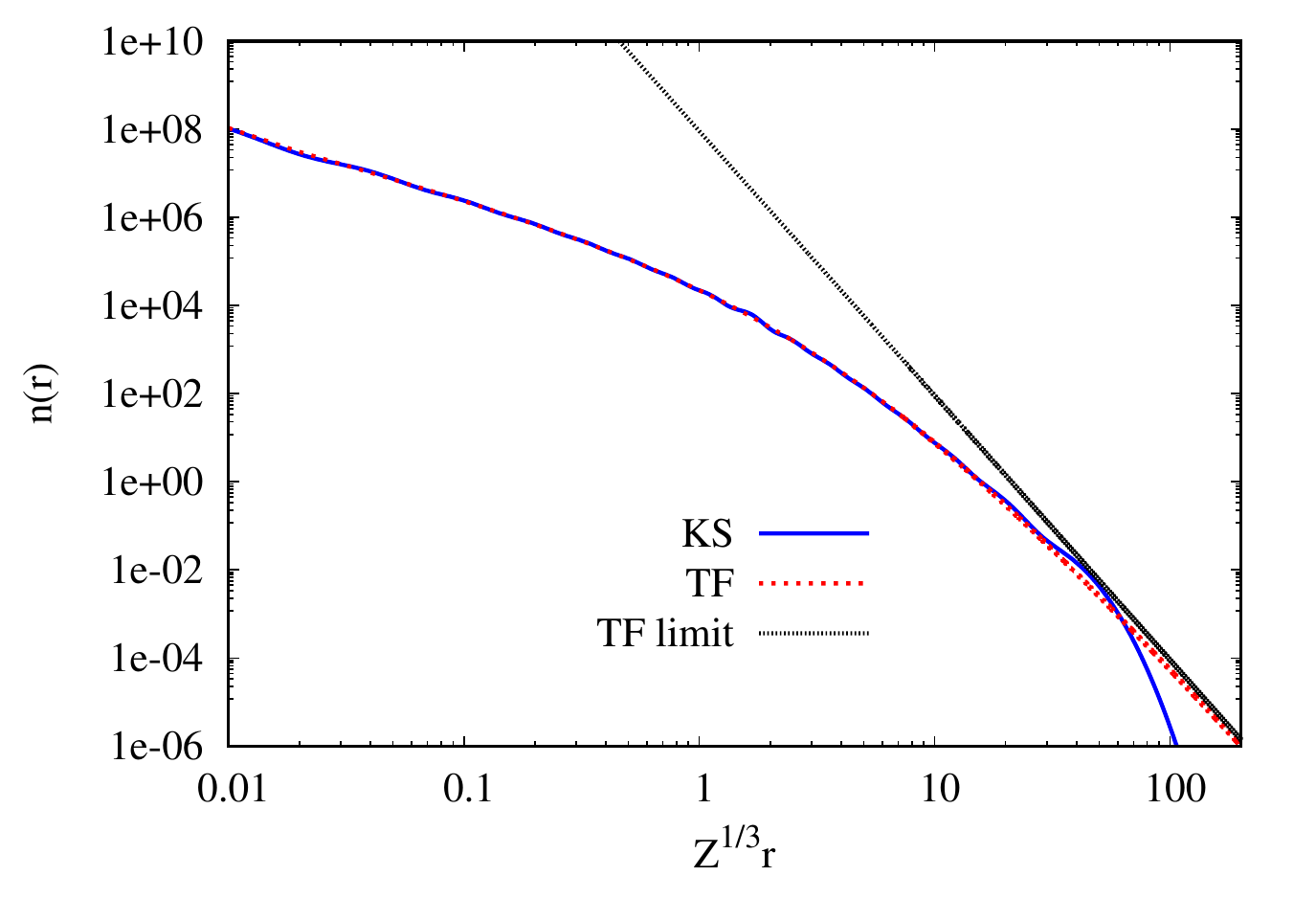}
		\caption{\label{fig:denstest}Comparison of KS density (KS), the TF density (TF), and the asymptotic TF limit (TF limit) for the Z=976 neutral atom.}
\end{figure}

As a partial confirmation of our method,
we compare densities generated by FHI98PP and the Aufbau principle for 
the mathematical element with $Z=976$ to the Thomas-Fermi density using the 
numerical parameterization of Ref.~\onlinecite{LCPB}.   
Fig~\ref{fig:denstest} shows the 
KS density (blue line), the TF density (red dotted line), and the TF limit of 
the density (black dotted line) versus scaled radius for element number 976.  
The scaling of $Z^{1/3}$ reflects the radius of the atom in the TF approximation, with peak radial density occurring for $Z^{1/3}r\sim 1$.
Note the high agreement between the KS density and the TF density over a large
range in scaled radius.
However, though suppressed by the log plot, shell structure is evident in the
KS density as oscillations around the TF density, wih the density deviating
from the TF limit especially for the last oscillation or two ($Z^{1/3}r>10$).  
As expected the density diverges from the TF density for very large
values of scaled radius, in the region of exponential decay beyond the last
occupied shell.  The Thomas-Fermi model assumes an infinite number of 
particles and continues indefinitely with the density decaying as $1/r^6$.  
The KS density never quite reaches the $1/r^6$ large-$r$ limit of the TF
density
(the ``mantle'' of the core of Ref.~\onlinecite{liebsimon}) and 
in this sense has not completely reached the TF limit. 

\subsection{Pauli Potential: near the nucleus \label{sec:nearnucleus}}

\begin{figure}[!htbp]
		\includegraphics[width=1.0\linewidth,height=0.41\textheight,keepaspectratio]{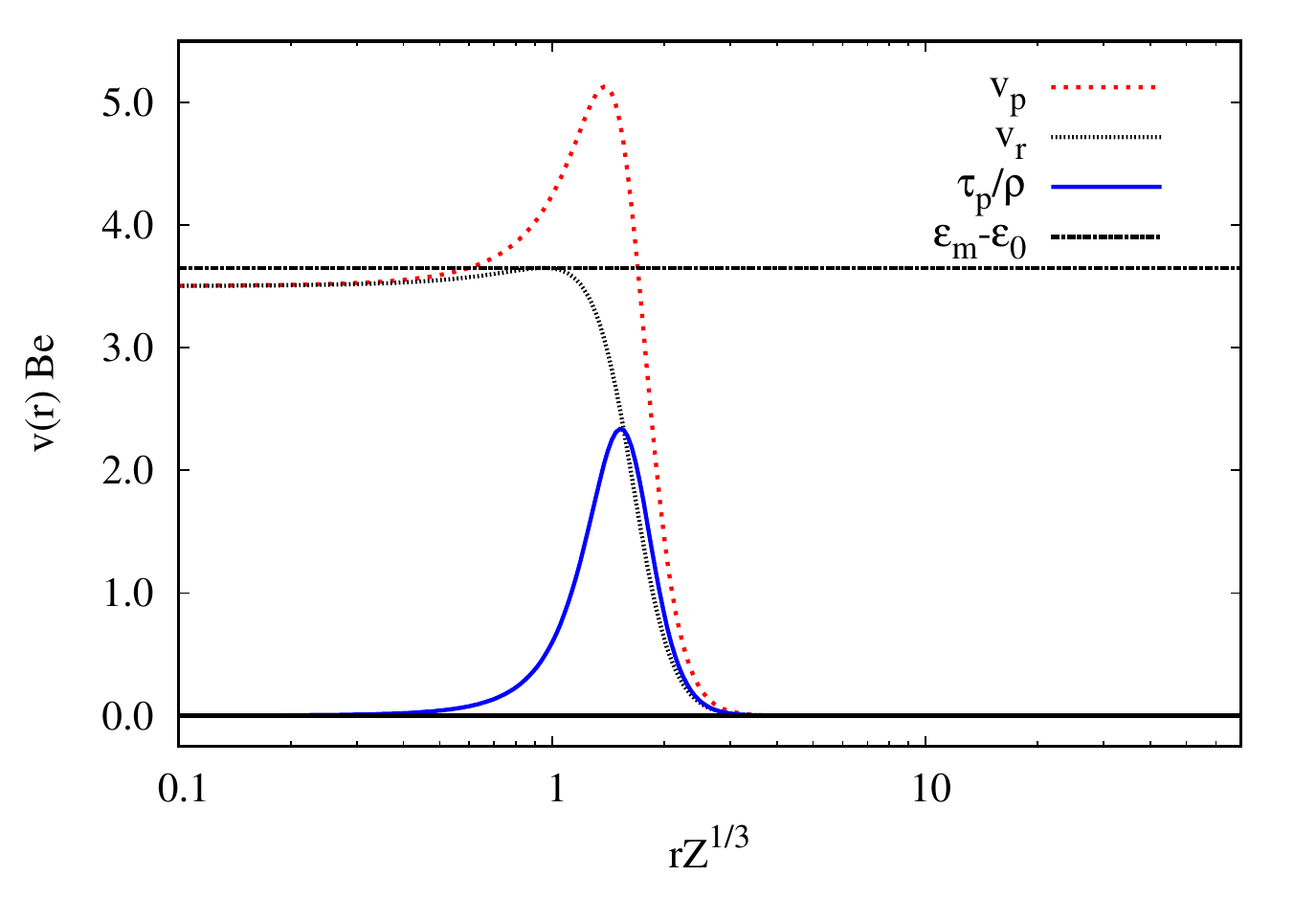}
		\caption{\label{fig:vpBe} Various contributions to the Pauli
   potential for Be.  Blue is the Pauli KED divided by the density, 
      close black dotted is the response potential, and red dotted is the Paul 
      potential.
      The eigenvalue difference $\epsilon_M-\epsilon_0$ between
      highest and lowest occupied orbitals is shown as a solid black line.
      }
\end{figure}

As suggested by Lieb and Simon's schema for describing the
large-$Z$ atom, it is helpful to investigate Pauli potentials for separate
regions of space.  We thus examine first, the potential of the one 
or two electron shells nearest the nucleus, then that of 
the core and valence shells, 
and finally the evanescent behavior far from the nucleus.
  Fig~\ref{fig:vpBe} shows the Pauli potential and the two components that are used to construct it via Eq.~(\ref{eq:vpauli}) -- the Pauli KED divided by the density and the response potential.  The constant $\epsilon_{M}-\epsilon_{0}$ is shown as a solid red line.

Beryllium is a usefully didactic system since it has only two shells -- it is in effect a two-state system and the simplest atomic structure that has a non-zero Pauli
contribution.  The Pauli KED is nonzero only in the transition region between the 1s and 2s shells.  Further out, it is zero because only the 2s shell effectively contributes to the KED -- it becomes effectively a single state system indistinguishable from the bosonic case.  Inside, the issue is more complicated. The 2s shell has a small nonzero piece and so naively one would expect the Pauli KED to be nonzero, but as we discuss in detail below, the Pauli KED is exactly zero at the nucleus as a consequence of the nuclear cusp condition on the density.

The response potential $v_r$ for Be is essentially a step function with a 
single step from the 1s shell, having an eigenvalue close to the hydrogenic 1s 
value, to the 2s shell. 
The second shell is the highest occupied energy shell and thus, given the 
definition of $v_r$ [Eq.~(\ref{eq:vrespexact})], makes zero contribution to 
the numerator of the response potential.  The response potential of the 
lowest energy shell agrees reasonably well near the nucleus by the two-state 
energy difference 
$\epsilon_M-\epsilon_0 = \epsilon_{2s} - \epsilon_{1s}.$\cite{baerends1997quantum}

The net effect on $v_p$ of the two contributions to it in Eq.~(\ref{eq:vpauli}) is instructive.  Recall the conceptual definition of $v_p$: 
given a system of fermions in an external potential that  one wishes
to replace with a fictitous system of bosons with the same ground state
density, then $v_p$ is the potential that one needs to add to the 
external potential in the bosonic system to achieve this.
Here the goal is to make the density from a single bosonic state 
$\psi \sim \sqrt{\rho}$ duplicate the two shells of the fermionic system.  
This is done by creating a potential step (due to $v_r$) that pushes density out of 
the 1s shell region into the 2s shell region, while an additional barrier (due 
to $\tau_p/\rho$) separates the charge into two distinct shells.  Finally, one 
may note that the values of $v_p$ and $v_r$ at contact with the nucleus 
are equal to each other and slightly less than $\epsilon_M-\epsilon_0$.
	
\begin{figure}[!htbp]
		\includegraphics[width=1.0\linewidth,height=0.41\textheight,keepaspectratio]{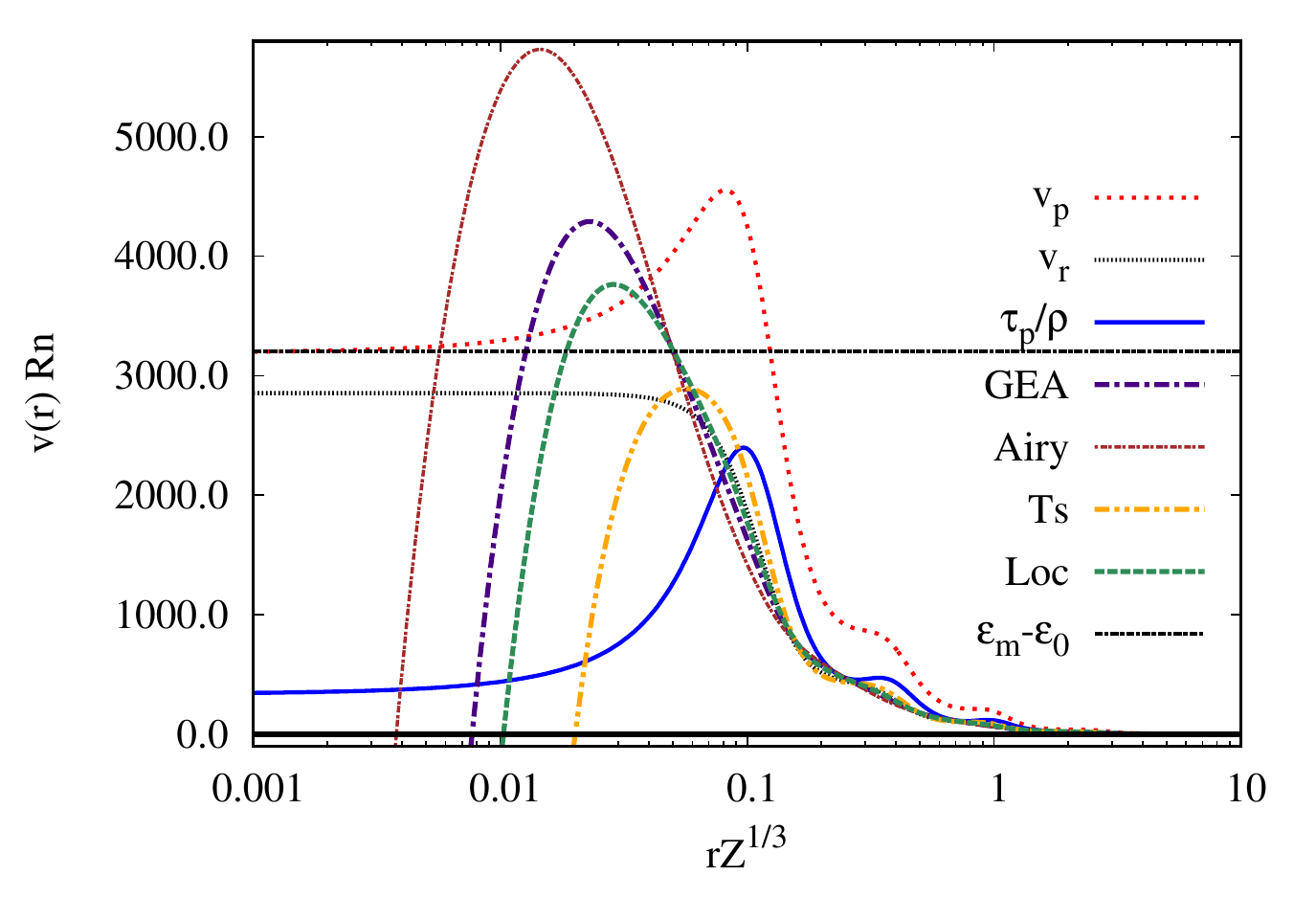}
		\caption{\label{fig:vpRn}
Components of the Pauli potential,
$\frac{\tau_p}{\rho(\bld{r})}$, $v_p$, and $v_r$, for Rn, compared to 
response potential from several DFT models.  
GEA is the gradient expansion approximation [Eq.~(\ref{eq:taugea2})], 
Airy, the Airy gas gradient expansion [Eq.~(\ref{eq:gealoc})], 
loc, the fit to the local KED for high-$Z$ atoms [Eq.~(\ref{eq:tauairy})], 
Ts, the Tsirelson model [Eq.~(\ref{eq:TSprefit})]. 
$\epsilon_M-\epsilon_0$ is difference between highest and lowest occupied eigenvalues. 
}
	\end{figure}

As there are known asymptotic behaviors for both total energy and near nuclear energy densities related with large Z scaling for the Kohn-Sham KE, 
it is of interest to analyze the large Z scaling of $v_p$.
Fig~\ref{fig:vpRn} plots the same quantities as Fig~\ref{fig:vpBe}, but for Rn.  In addition we plot three gradient expansion models for the response potential discussed Sec~\ref{Theory} --  the canonical GEA (purple dashed),  the fit to the KED of high-Z atoms ($v_{Loc}$) (green dash), and the model of Ref.~\onlinecite{Ts} (yellow dashed).  Note that every potential visually has a three step structure with transitions at $Z^{1/3}r=0.1$ and $Z^{1/3}r=1$, related to the three innermost of six occupied shells (the remaining three shells are too small to see in this plot).  In comparison to Be, $v_r$ seems to retain the step structure and $\tau_p$ has weak local maxima in between shells, but the shell structure overall is less pronounced.  

Note that for Rn, $v_p$ is almost exactly $\epsilon_m-\epsilon_0$ in the 
near-nuclear region.  
This trend continues to improve as $Z\to\infty$, however visually Rn 
essentially shows complete agreement between $\epsilon_m-\epsilon_0$ and $v_p(r=0)$, so no larger Z atoms are plotted in this fashion.  
The actual contact value for $v_p(0)$ is much larger than that for
Be -- the energy scale is roughly $Z^2$, that of the noninteracting 
hydrogen-like system.
At the same time $\tau_p(0)$ is definitely non-zero and so $v_r(0)$ and 
$v_p(0)$ are no longer the same value.

The GEA models all trend toward $-\infty$ as $\bld{r}\to0$.  This is due to the charge singularity at the origin, resulting in a Laplacian of the density that 
diverges in this limit.  
This is a flaw in any GEA model
and is caused by the divergence term in Eq.~(\ref{eq:vksapprox}).
It is notable that the TS model does come close to predicting the turning point for $v_r$.  

\subsubsection{Analytic analysis of nuclear region}
	
	It seems from Fig.~\ref{fig:vpRn} that as $Z\to\infty$, the value of 
$v_p$ near the nucleus approaches a constant equal to 
$\epsilon_M-\epsilon_0$. At the same time, $\tau_p/\rho$ does not seem to 
converge to zero here, but rather to a value about 10$\%$ of $v_p(0)$; 
thus $v_{r}(0)$ falls short of $v_p(0)$ by the same amount.
These asymptotic behaviors can be proven mathematically. 
	
Naively, if one considers Eq.~\ref{eq:vrespexact} 
as $\bld{r}\to0$, and assume that the 1s orbital is the primary contribution
to this equation, one gets
	\beq
		\label{eq:vrespzeroapprox}
		\lim_{r\to0}v_{r}\approx(\epsilon_M-\epsilon_0).
	\eeq
	This can not be exactly true because every s orbital has a contribution at the nucleus.  

To improve the description, we consider what happens near the nucleus in the 
limit that nuclear charge $Z$ and electron number both go to infinity.
Even for finite $Z$ the effect of electron-electron interactions
becomes very small compared to the nuclear potential and thus
they can be ignored.  Low-lying energy eigenvalues approach in energy and 
degeneracy those of the corresponding noninteracting system --
a Coulombic nuclear potential with charge $Z$.
One thus can use hydrogenic wavefunctions to construct both $\tau_p$ and
$v_{r}$ near the nucleus.  This should be accurate out to a radius of 
order $r/Z$ where Lieb and Simon~\cite{liebsimon}  show that the electron density starts to resemble that of the Thomas-Fermi atom.

To build a model for $v_r$ based on this picture we first note that
for a hydrogenic central potential, the value at the nucleus of the radial 
component $R_{nl}$ of an eigenfunction is
\beq
		\label{eq:Rhydend}
		R_{nl}(0) = \sqrt{4\left(\frac{Z}{n}\right)^3} \delta_{l0}. 
\eeq
Here the identity $L_{n-1}^{1}(0)=n$ for Laguerre polynomials has been used.
Next, we apply this result to the exact expression for $v_{r}$
[Eq.~(\ref{eq:vrespexact})] by defining the net density of an angular 
momentum subshell,
\beq
      \rho_{nl}(r) = \frac{(2l + 1)}{4\pi} \left|R_{nl}(r)\right|^2
      \label{eq:rhonl}
\eeq
and, following Ref.~\onlinecite{heilmannlieb}, the density of a complete 
energy shell:
\beq
      \rho_{n}(r) = \sum_{l=0}^{n-1} \rho_{nl}(r).
\eeq
For a hydrogenic system, given the degeneracy in energy over angular
momentum quantum number $l$, one has
\beq
      v_{r}(r) = \sum_{n=1}^M 
          \left( \epsilon_M - \epsilon_n \right) \rho_n(r)/\rho(r).
      \label{eq:vresphydrogen}
\eeq

Since we are considering the limit $Z\to\infty$, it is appropriate also to 
take the limit that $M\to\infty$, that is, the perfect ``Bohr atom'' where
all orbitals are filled and all are given by those of the noninteracting
hydrogen atom.  
At the origin, only $l=0$ contributes, so $\rho_{n}(0) = \rho_{n0}(0)$.
Thus Eq.~(\ref{eq:vresphydrogen}) reduces to
	\beq
		\label{eq:vrespzero}
		v_{r}(0) = \braket{\epsilon_{n0}}
              = {\displaystyle -\frac{\sum_{n=1}^{\infty} 
                     \epsilon_{n0} \rho_{n0}(0)}
                     {\sum_{n=1}^{\infty} \rho_{n0}(0)}},
	\eeq
where we assume $\epsilon_{\infty}=0$.
Substituting in Eqs.~\ref{eq:rhonl} and~\ref{eq:Rhydend} and summing over 
$n$ gives
	\beq
		\label{eq:rhohydrogen}
		\rho(0)=\sum_{n=1}^{\infty}\frac{2}{\pi}\left(\frac{Z}{n}\right)^{3}.
	\eeq
Similarly, we use $\epsilon_n = -Z^2/2n^2$ for a hydrogenic atom and 
repeat this process to get the numerator of Eq.~\ref{eq:vrespzero}.
With a bit of manipulation
one can write the ratio as
	\beq
		\label{eq:vrespzerolargeZ}
		 {\displaystyle 
        		v_{r}(0) = {\frac{Z^2}{2} \frac{ \zeta(5)}{ \zeta(3)},
        		}
    	}
    	\eeq
    	where
    	\beq
    		\zeta(s)=\sum _{n=1}^{\infty }{\frac {1}{n^{s}}}
    	\eeq
    	is the Riemann-Zeta function.  
  Given that $\zeta(5)=1.03692$ and $\zeta(3)=1.20205$, one has as $Z\to\infty$
    	\beq
    		v_{r}(0)=0.862626\frac{Z^2}{2}=-0.862626\epsilon_0.
	\eeq
    	
The Pauli KED near the nucleus can be analyzed in a similar fashion, with
more difficulty, since it necessarily involves derivatives of orbitals.
It is fairly straightforward to show that the contribution of $s$-orbitals
to the KS KED exactly equals the von Weizsacker KED in this region.  
In effect, this
describes the connection between the cusp conditions near the nucleus 
obeyed by $s$-orbitals, and that of the total density.  Somewhat 
counterintuitively, $p$-orbitals also have a nonzero contribution to the KS KED
near the nucleus, both radially and from their nonzero angular momentum\cite{cancioredd, thesis, acharya, constantin2016kinetic}.  
It is the contribution from these orbitals that cause $\tau_p$ to be nonzero
near the nucleus.

The Bohr atom model used here for $v_{r}$  has recently been used by 
Constantin et al. to analyze the large-$Z$ limit of $\tau_p$ at the nucleus. 
In this case, they show (Eq.~(20) of Ref.~\onlinecite{constantin2016kinetic})
    	\begin{eqnarray}
    		\label{eq:taupconst}
    		\lim_{Z\to\infty} \tau_p(0)&=& 
                         \sum_{n=1}^{\infty} 3\tau_{vW}[\rho_{n1}](0) \\
                         &=& \sum_{n=1}^{\infty} \frac{(n^2 - 1)Z^5}{\pi n^5},
	\end{eqnarray}
where $\tau_{vW}[\rho_{nl}]$ is the vW KED evaluated using the density of 
the $(nl)$ angular momentum subshell.
The end result is closely related to that for the $Z\to\infty$ limit of 
$v_{r}(0)$:
	\beq
		\label{eq:taupoverrho}
		\frac{\tau_p(0)}{\rho(0)} = 
                     \frac{Z^2}{2} \frac{\zeta(3)-\zeta(5)}{\zeta(3)}
	\eeq
	and therefore, using Eq.~(\ref{eq:vpauli})
	\beq
		\label{eq:vpepsilonnot}
		\lim_{Z\to\infty} v_p(0)= \frac{Z^2}{2}
	\eeq

This may be recast in a form that is more robust as well as conceptually
revealing.  First we note that the limit $Z^2/2$ is shared with the 
lowest orbital eigenvalue:
       \beq
          \lim_{Z\to\infty} \epsilon_0 = -\frac{Z^2}{2}.
       \eeq
Then, observing that the form of the response potential 
involves a difference between the highest occupied eigenvalue 
$\epsilon_M$ and the other occupied eigenvalues,
and noting that $\epsilon_M$ is a small energy independent of $Z$,
we posit the general limit for $v_p(0)$:
	\beq
		\label{eq:vpepsilon}
		v_p(0)\sim \epsilon_M-\epsilon_0.
	\eeq

\begin{figure}[!htbp]\centering
		\includegraphics[width=1.0\linewidth,height=0.41\textheight,keepaspectratio]{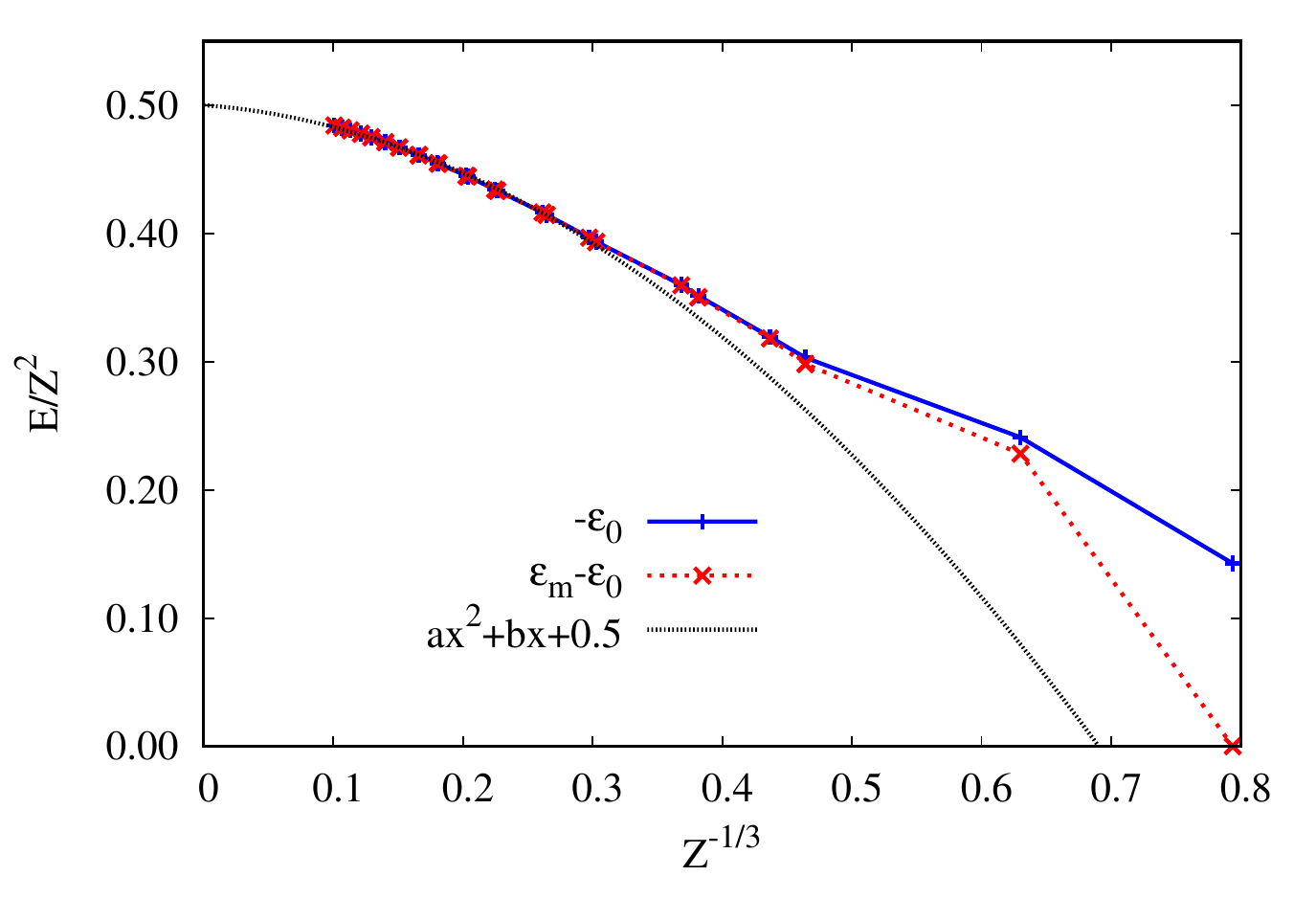}
		\caption{\label{fig:vpZenot}
Analysis of lowest energy eigenvalue of large-$Z$ atoms 
versus the small parameter
$Z^{-1/3}$ for the large-$Z$ expansion of atomic energies.
We show $-\epsilon_0+\epsilon_M$, $-\epsilon_0$, and a curve fit of the 
form $ax^2+bx+0.5$ versus 
$x=Z^{-1/3}$ for noble gases from He to $Z=976$. }
	\end{figure}

We verify these assumptions first by plotting, in
Fig.~\ref{fig:vpZenot}, ($\epsilon_M-\epsilon_0)/Z^2$ (red dashed) 
and  $-\epsilon_0/Z^2$ (blue) for alkali metals and noble gases from 
He to $Z=976$.  
These are plotted against the small parameter $Z^{-1/3}$ 
that characterizes the large-$Z$ expansion of atomic energies.
We fit this trend with a polynomial form $(ax^2+bx+0.5)$ with $x = Z^{-1/3}$.
A least squares regression results in $a\!=\!-0.879 \pm 0.017$ and $b\!=\!-0.091 \pm 0.004$. 
The fit is highly accurate for large Z atoms, starts to deviate from the
observed around $Z<64$, but is still within 10\% of the true value for Ne.
One may note that dropping $\epsilon_M$ from the approximation for $v_p(0)$ 
affects the result primarily for He where in fact $\epsilon_M=\epsilon_0$.
At the same time, the value of $-\epsilon_0$ is significantly 
off the hydrogenic value of 0.5 for any realistic value of $Z$.

\begin{figure}[!htbp]\centering
		\includegraphics[width=1.0\linewidth,height=0.41\textheight,keepaspectratio]{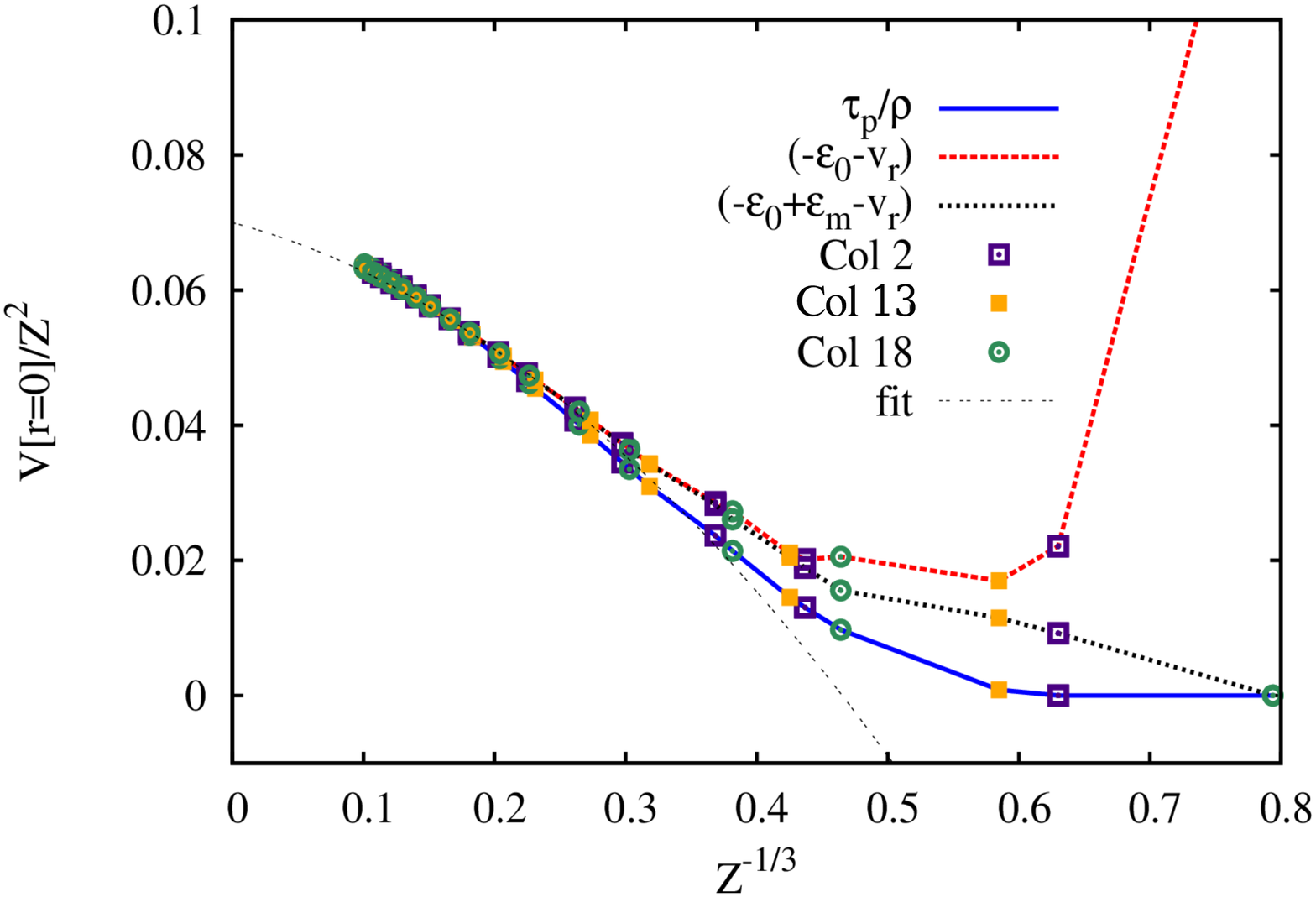}
		\caption{\label{fig:vpZ}Contact values for $\tau_p/\rho$,~$-\epsilon_0-v_{r}$, and $\epsilon_M-\epsilon_0-v_{r}$ as a function of $Z^{-1/3}$.  Values taken atoms from columns 2 and 13 and 18 respectively, 
extended to $Z=976$, and compared to a parabolic fit $cx^2+dx+0.07$.
}
	\end{figure}

To test our assumptions of the finite-$Z$ value
 of $v_p(0)$ [Eqs.~(\ref{eq:vpepsilonnot}) and (\ref{eq:vpepsilon})] 
work,  
we next plot in Fig.~\ref{fig:vpZ} the value of 
$v_p - v_r$ (blue) at the origin for all atoms in columns 2, 13, and 18 of
the periodic table, extended to $n=16$ ($Z=976$.)  
This is again scaled by $Z^2$ and plotted versus $Z^{-1/3}$. Subtracting 
off $v_r$ removes the large majority of the Pauli potential at the origin,
leaving a relatively small piece (equal to $\tau_p/\rho$) which makes 
the error in our limiting ansatz readily visible.
Then we compare to the difference between the large $Z$ 
limit $\epsilon_M-\epsilon_0$ and $v_r$ (black dashed line) and repeat for the
the less accurate limit $-\epsilon_0$ (red dashed).
Data is taken from three columns of the periodic table, and differentiated
by plotting points of different types.
A complete table of data used to generate fig.~\ref{fig:vpZenot} and fig.~\ref{fig:vpZ} /is included in supplemental materials.

Note that all three functions of Z approach the same limit as $Z\to \infty$,
converging to less than 1\% error in $v_p(0)$ by roughly $Z=36$.  
Furthermore this dependence is column independent-- curves from each column plotted fall onto the same trend after just one shell.  This makes sense since we are measuring the Pauli
potential at the nucleus, where presumably the effects of a variably filled
valence shell should be minimal.
The effect of including $\epsilon_M$ in our model is felt most for 
single-shell systems like He, 
where it retrieves the exact value of zero for $V_p(0)$.  
We make a parabolic fit of the data to the trend 
$cx^2+dx+0.07$ where $x=Z^{1/3}$ (black dash-dotted line).
A least-squares regression of our data at large $Z$ 
results in $c=-0.223 \pm 0.009$ and $d=-0.0491 \pm 0.0024$.  
The value of 0.07 for the $y$ intercept is determined using 
Eq.~\ref{eq:taupoverrho}. 
The fit has a very weak linear term, indicating
that the contact value of $v_p$ roughly varies with nuclear charge 
as  $0.5 Z^2 + cZ^{4/3}$.
 
	\subsection{Core and Valence}
	
Next we consider the behavior of the Pauli potential
and its constituents $\tau_p/\rho$ and $v_{r}$, away from the nucleus.
For the ease of visualization across many shells, 
we employ unitless, scale-invariant quantities.   For the kinetic energy
density, it is common to do so by defining an enhancement factor, $F$, relative 
to the Thomas-Fermi KED:
       \beq
          F = \tau_{KS} / \tau_{TF}
       \eeq
and equivalently, a Pauli enhancement factor defined by
       \beq
          \tau_{p} = F_p\tau_{TF}
       \eeq
so that $F_p$ is given by
       \beq
	\label{eq:enhancement}
	F_p = \left(\tau-\tau_{vW}\right)/\tau_{TF}.
       \eeq
For any model for the Kohn-Sham kinetic energy density, 
model Pauli enhancement factors may be similarly defined.
Beyond the obvious advantages of scale invariance, the Pauli enhancement
factor for the KS KED is closely related to the Electron Localization Factor (ELF)~\cite{BeckeEdgecombe} and is equal to the $\alpha$ term used in 
meta-GGA functionals.~\cite{Becke98,SXR12} 
In order to produce a unitless representation of the 
Pauli potential and its components, we scale each quantity by 
$\tau_{TF}/\rho$, the ratio of KE and particle densities in the 
TF model.  This is $3/5$ of the local fermi energy $\epsilon_F$ in the 
TF picture.

\begin{figure}[!htbp]\centering				    			  
                   \subfigure[]{
                   \includegraphics[width=1.0\linewidth,height=0.41\textheight,keepaspectratio]{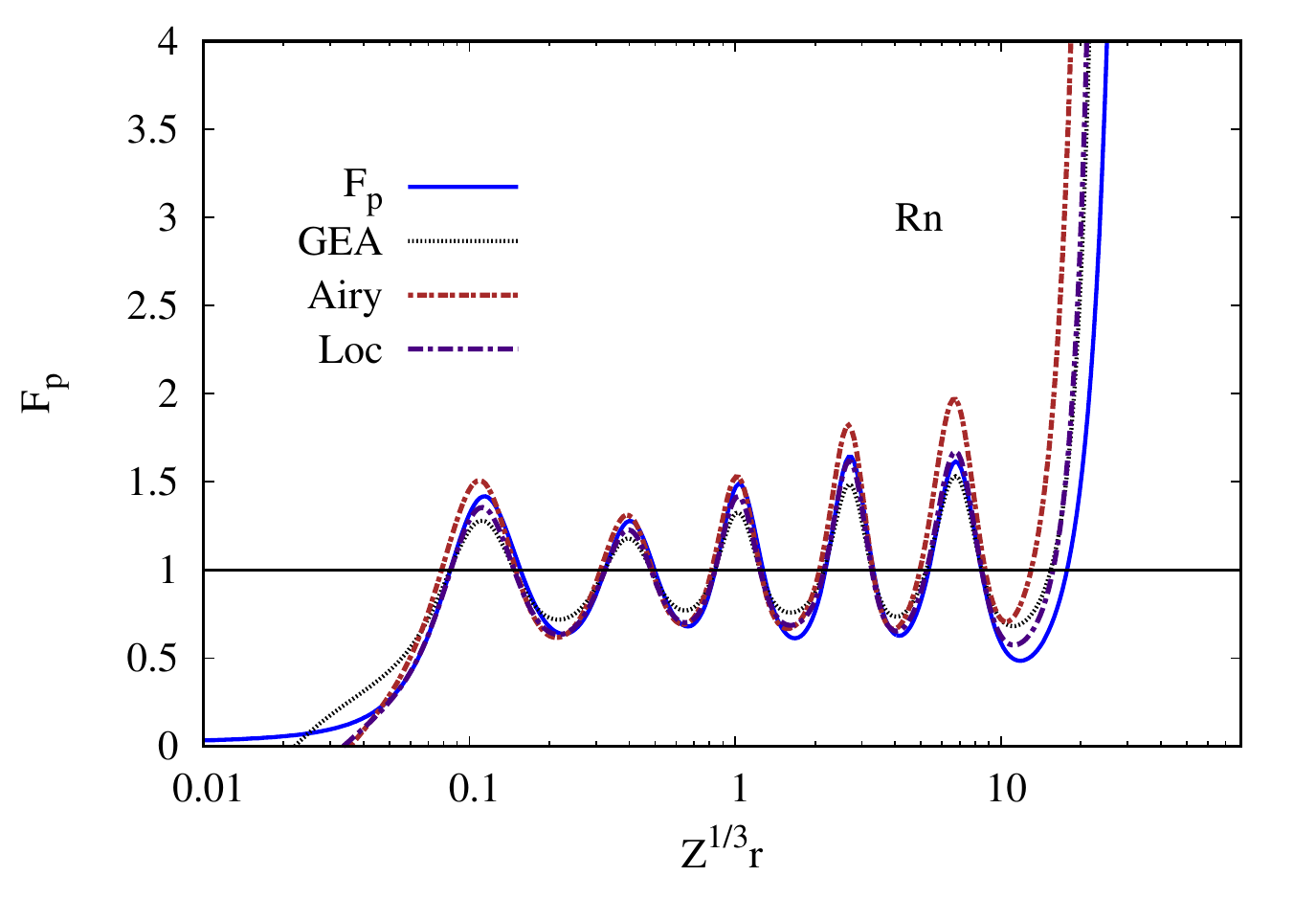}
                   }
                   \subfigure[]{
\includegraphics[width=1.0\linewidth,height=0.41\textheight,keepaspectratio]{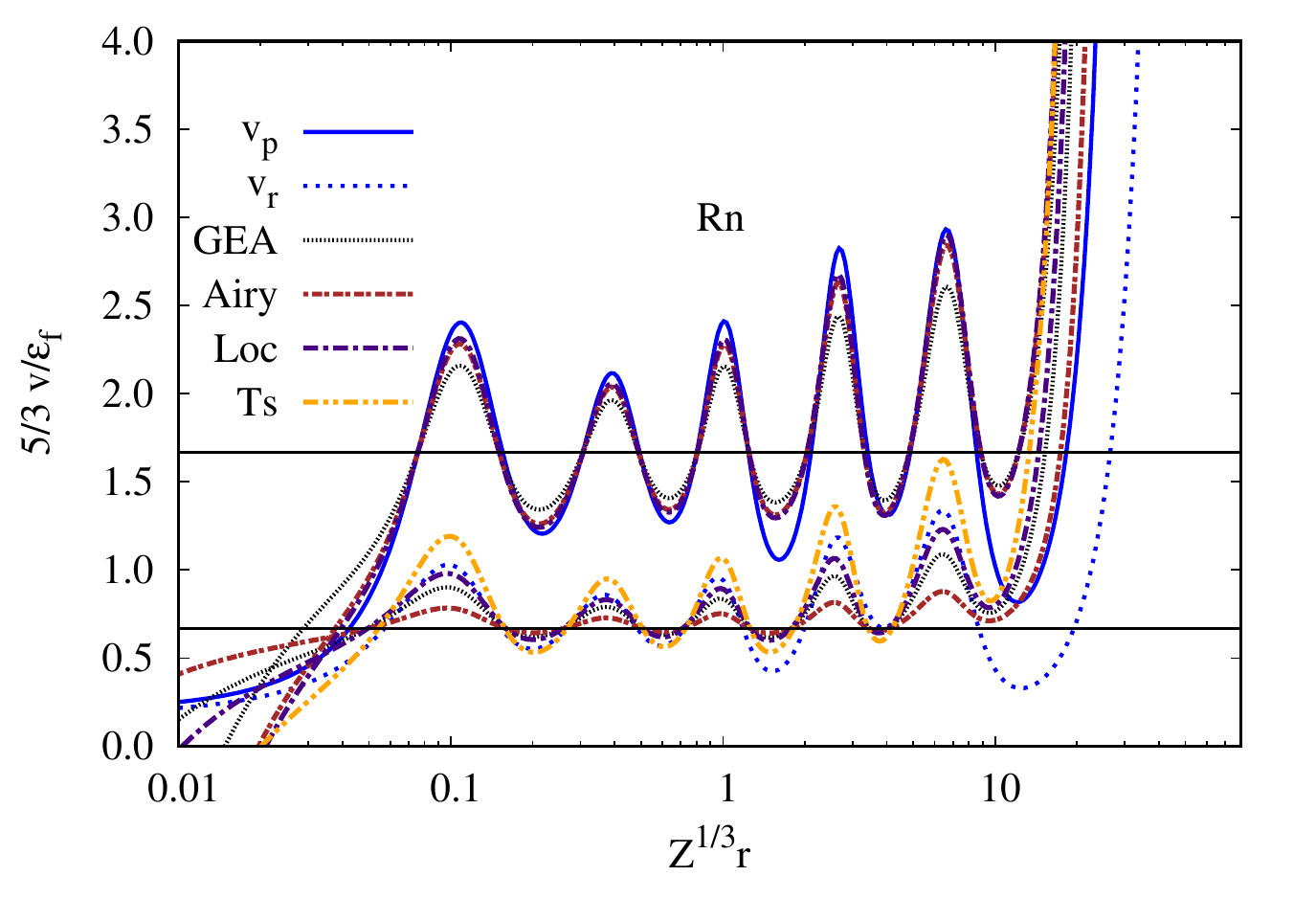}
                   }
		\caption{\label{fig:Fresp_Rn}
Unitless representation of KE potentials for radon.  
(a) Pauli enhancement factor $F_p$ for the Kohn Sham KED, compared to 
several variations of the gradient expansion approximation, versus scaled 
radius $Z^{1/3}r$.  Models shown are the standard gradient expaonsion (GEA),
Airy gas model (Airy), fit to the local KED of atoms (Loc) and the Tsirelson
model (Ts);
the Thomas-Fermi limit is shown as the solid horizontal line.
(b) Pauli potential $v_p$ (solid blue line), response potential $v_r$ (dashed blue line), and gradient expansion approximations of the same, 
scaled by $\tau_P/\rho$.
Each model is shown with the same color and dashing for 
$v_p$ and $v_r$ but center on different TF limiting cases -- 5/3 for the
former, 2/3 for the latter.
}
	\end{figure}

Fig.~\ref{fig:Fresp_Rn}(a) plots 
the Pauli enhancement factor $F_p$ for the Kohn-Sham KE density of radon.
This is compared to various GE approximations: 
the standard gradient expansion, the Airy gas model~\cite{lindmaa14}, the local fit to the gradient expansion of Ref.~\onlinecite{cancioredd} and the model of 
Ref.~\onlinecite{Ts}.  These are plotted against the scaled distance 
$x = Z^{1/3}r$, chosen so that the peak radial probability density in the 
TF model for any atom occurs at roughly $x\!=\!1$.
The constant line at one shows the TF limit for $F_p$. 

A notable feature of these plots is the nearly periodic oscillation 
of the exact KE density and the GE models about the TF limit.  
This behavior reflects the shell structure of the atom: 
a value of $F_p < 1$ indicates a region dominated by a single shell, 
producing a value for $\tau_{KS}$ lower than that predicted by TF theory, while
the opposite is true for $F_p > 1$.
Thus each minimum indicates a different principal quantum 
shell.  
The five maxima show the regions of transition between the six shells of Rn, 
while the last exponentially divergent tail at large $r$ is the classically 
forbidden evanescent region outside the atom.
It is interesting the oscillations have a roughly equal period in a 
semi-log plot, suggesting exponential growth in the period of quantum
oscillations.  

All versions of the gradient expansion recover the main qualitative features 
of the Pauli enhancement factor away from the nucleus and for the most part
are quite accurate quantitatively.
The quantum oscillations of the Tsirelson model start to deviate from the 
TF limit in the outer three shells; also the Airy gas model 
overestimates the true enhancement factor by a scaling factor of roughly 10/9.  

Fig.~\ref{fig:Fresp_Rn}(b) plots unitless representations of the Pauli potential $\rho v_p/\tau_{TF}$ and response potential $\rho v_r/\tau_{TF}$ versus
scaled radial distance $Z^{1/3}r$ for radon.  These
are compared to various gradient expansion models, as before.
The Pauli potential is not plotted for the Tsirelson model 
because of its dependence on the functional derivative of other quantities
like exchange and correlation.  
Referring to Eqs.~(\ref{eq:vpauliGE}) and~(\ref{eq:vrespGE}), we see that
the Thomas-Fermi limit of the scaled Pauli potential is the constant 5/3, and 
that of the response potential, 2/3, shown as black horizontal lines. 

Note that the scaled potentials show the same shell structure as the Pauli
enhancement factor, oscillating about their TF limis with peaks and minima 
in nearly the same locations.  The center of the oscillations starts 
to deviate from the TF line slightly for the outer three shells.
This trend is better matched by the ``non-canonical'' GE's like the Loc and Airy models than by the standard GEA.
	
Comparing gradient expansion models for the Pauli potential 
[Fig.~\ref{fig:Fresp_Rn}(b)] we note that the Loc and Airy gas models
outperform the conventional GE over most of the atom.  Notably, the gradient
expansion of the Pauli potential [Eq.~(\ref{eq:vpauliGE})] only depends upon
the coefficient $\eta_P$ for the contribution of the gradient expansion 
from the gradient variable $p[\rho(r)]$.  Thus this data supports the 
use of a negative $\eta_P$ coefficient, as in the Airy gas and Loc models,
in contrast to the positive coefficient of the standard gradient expansion.
In addition the \textit{a priori} Airy gas potential 
is almost as good as the empirically fit Loc model.
At the same time, only the Loc model provides a close fit for the separate
pieces of the Pauli potential, $\tau_p/\rho$ and $v_r$  (the Airy gas
model is particularly poor for $v_r$, which underestimates the size of quantum 
oscillations by a factor of 3).   Thus Loc has the
best description of the coefficient $\eta_Q$ of the Laplacian term of the 
expansion. 

\begin{figure}[!htbp]\centering				    			  
                   \includegraphics[width=1.0\linewidth,height=0.41\textheight,keepaspectratio]{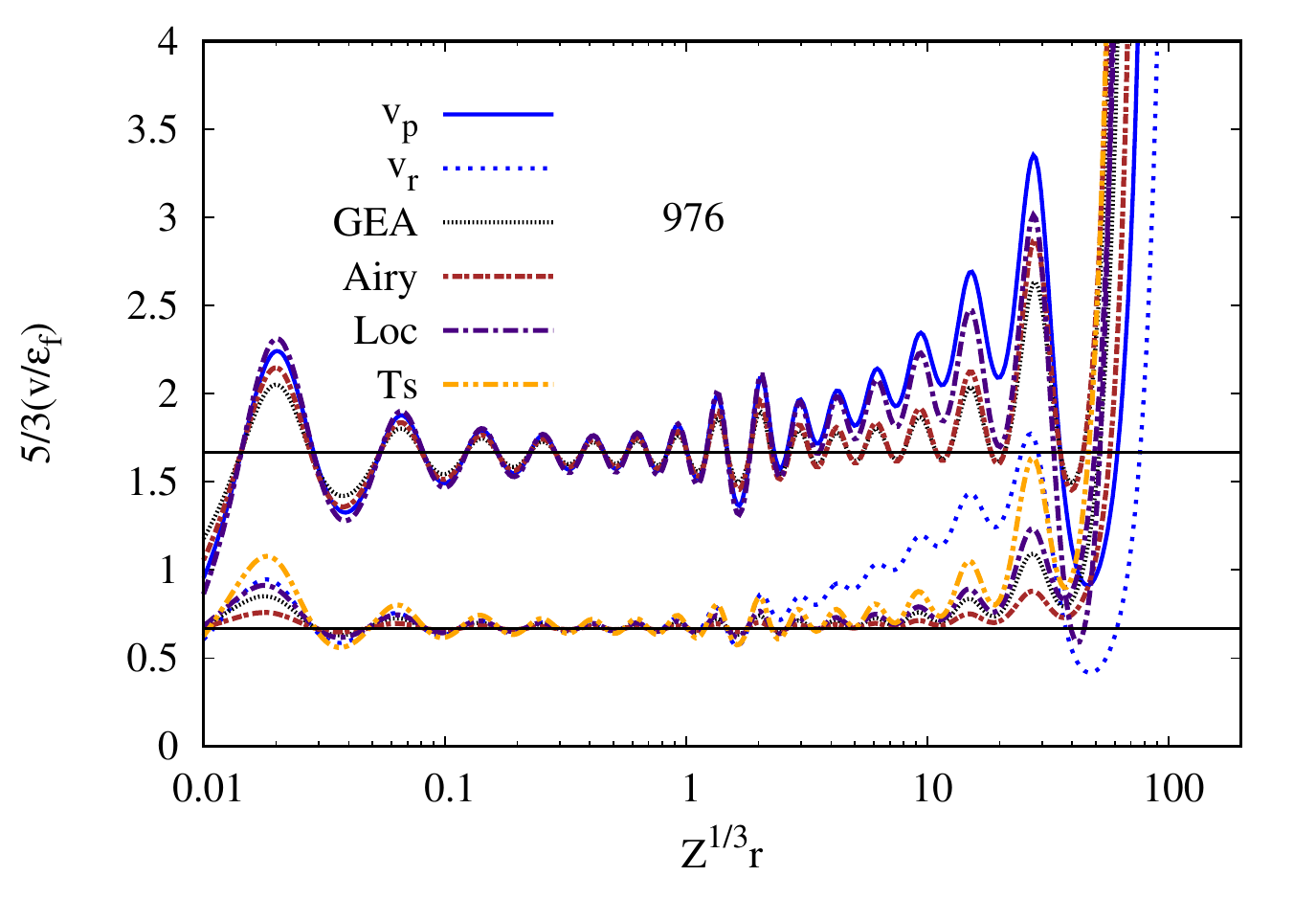}
          \caption{\label{fig:Fresp_976} 
Unitless representation of the Pauli potential $v_p$, response potential $v_r$, gradient expansion approximation of the response potential $GEA$, Airy gas response potential $Airy$, the local variation of the GEA response potential $loc$, and the Tsirelson response potential $Ts$ for element 976.}
	\end{figure}

We now consider the 
trends in atomic data as $Z$ is taken to be as large as practical in order
to try to piece out the high-$Z$ limit.
Fig~\ref{fig:Fresp_976}, similarly to Fig~\ref{fig:Fresp_Rn} (b), plots the unitless representation of $v_p$ and $v_r$, as compared to various GE 
approximations, for element 976.  This is the ``noble gas'' for row 16 of the periodic table mathematically extended using
the Aufbau principle -- thus there are 16 oscillations and 16 shells. 
Gratifyingly, these oscillations have considerably less amplitude than for radon, indicating passage towards the high-$Z$ limit.  Note that although the inner 
eight shells of $v_p$ and $v_r$ oscillate about the TF limit,
there is now an unmistakable trend away from the TF limit in the outer shells.  
This deviation is markedly absent in all the gradient
expansion models.  The deviation seems to be linear, and does not start 
until the middle shell of the atom is reached, around $Z^{1/3}r=1$.
The last oscillation in the potential, demarcating the valence shell,
occurs at $Z^{1/3}r\approx40$.  
Similar plots are made for the Column 2, 10 and 13 atoms from row 16 in 
the Supplemental Material.
		
\begin{figure}[!htbp]\centering				    			  
	\includegraphics[width=1.0\linewidth,height=0.41\textheight,keepaspectratio]{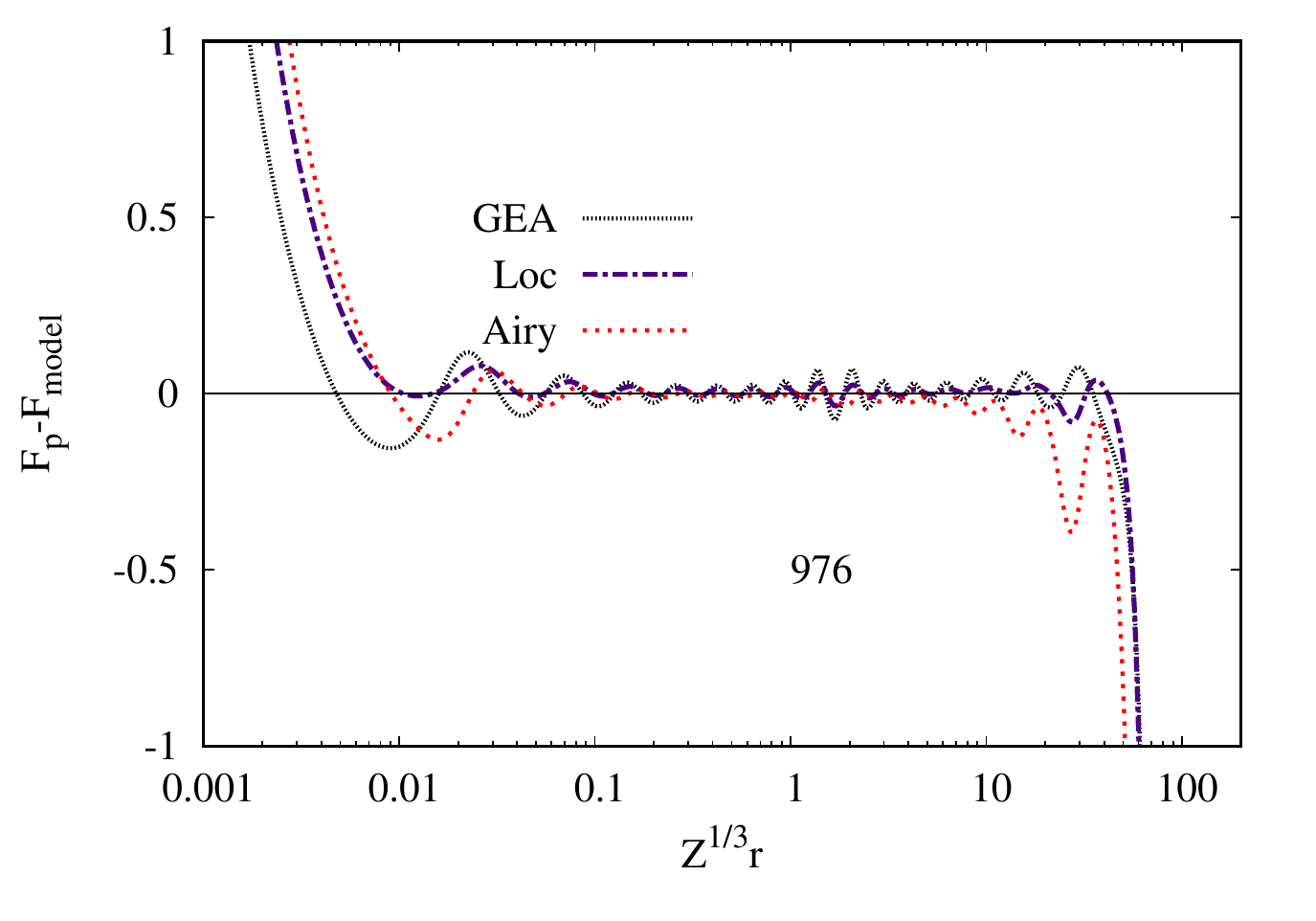}
	\caption{\label{fig:Fresp_976_error} Difference between the exact 
Pauli enhancement factor $F_p$ and various GEA models for element 976, as 
function of scaled radius.}
\end{figure}

To investigate further this unexpected behavior, we show 
in Fig.~\ref{fig:Fresp_976_error} the error in 
the Pauli enhancement factor $F_p^{model} - F_p^{exact}$ for the various
GEA models, for the $Z=976$ atom. This  highlights the 
``non asymptotic'' piece in $F_p$ -- the part that has not yet converged
to the Thomas-Fermi asymptote.  
The scaled Pauli energy density $F_p$ shows no unexpected behavior 
whatsoever.  The non-asymptotic remnant 
 can be fit extremely well by a zero line, and its amplitude of oscillation
is quite small.  This is consistent with the Lieb-Simon theorem
that the kinetic energy, obtained by integrating over the KE density must
tend to the Thomas-Fermi limit for large $Z$.  

\begin{figure}[!htbp]\centering				    			  
		\includegraphics[width=1.0\linewidth,height=0.41\textheight,keepaspectratio]{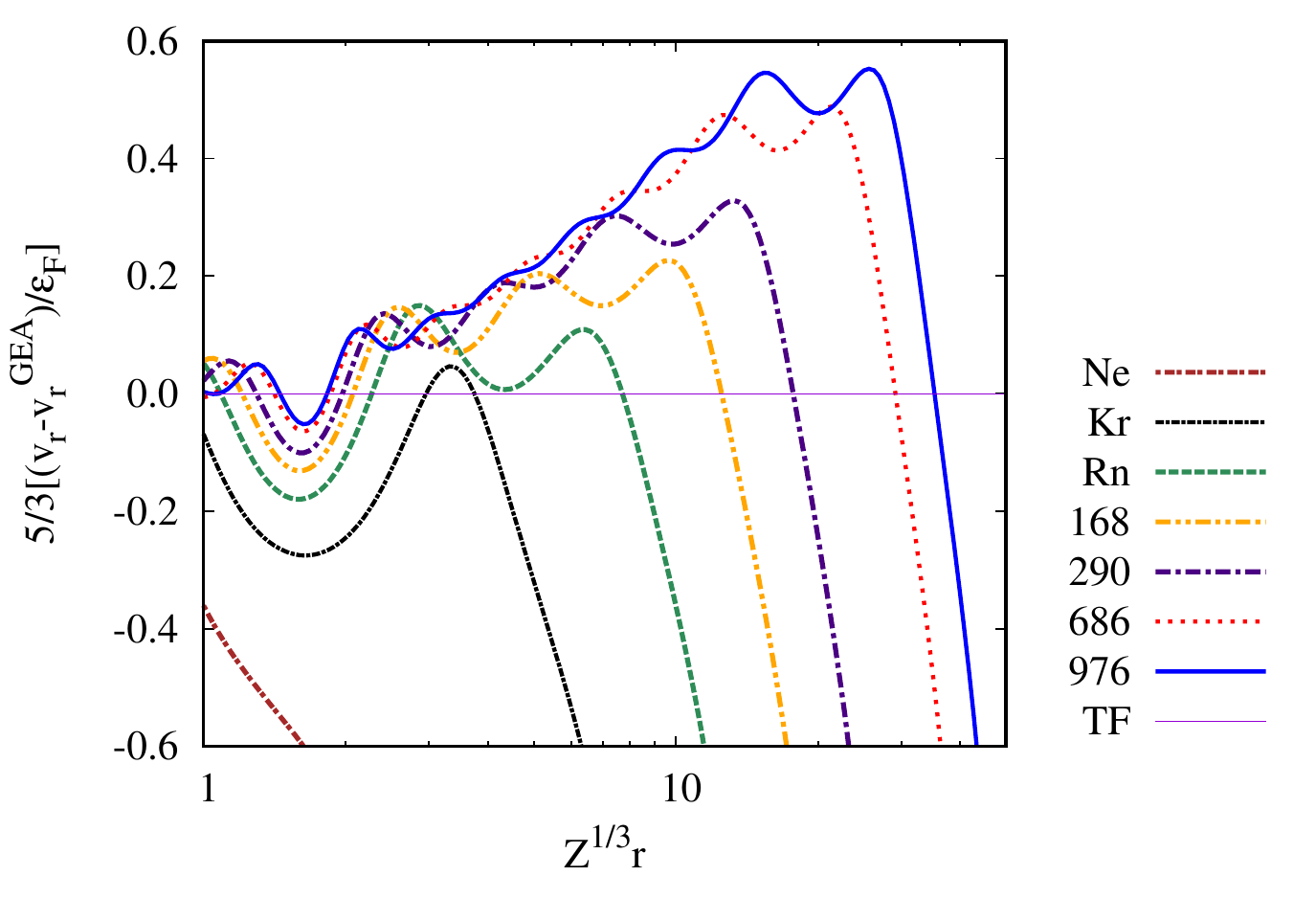}
		\caption{\label{fig:Fresp_pauli_row} Scaled difference between $v_r$ and GEA response potential for noble gases with even principle quantum numbers up to $n=16$.  The Thomas Fermi limit (TF) is narrow purple line; Ne, brown dashed; Kr, black dashed; Rn, green dashed; element 168, gold dot-dashed; 290, purple dot-dashed; 460, black dotted; 686, red dotted; and 976, blue solid.  
}
	\end{figure}

In contrast, 
Fig.~\ref{fig:Fresp_pauli_row} plots the difference between the exact response potential and the standard GEA approximation [Eq.~(\ref{eq:taugea2})] for noble
gas elements with even principle quantum number $n$ up to $Z=976$.  
The odd rows left out show a similar trend but with oscillations 
out of phase with the those of even rows.
The deviation from the TF limit seen in 
Fig.~\ref{fig:Fresp_976} here forms part of a trend, 
apparent at radon and growing consistently with $Z$, 
along a linear trendline versus $\log(Z^{1/3}r)$, 
from about $Z^{1/3}r \!=\! 2$ out to the edge of each atom.  
As $Z$ increases, the potential does not converge to the Thomas-Fermi
limit -- rather it deviates further away from it along this limiting
trend.
The valence shell (the last dip and peak before each curve drops to negative
infinity) deviates from the trend of the inner shells.  However it forms its 
own predictable linear trend away from the GE prediction, starting perhaps
with Kr, with same slope as the inner shells.

The difference between the exact response potential and the standard GEA model
for element 976 was fitted to a form $y = a\log(x/x_0)$ for $x\!=\!Z^{1/3}r$
and $y\! = \!\rho v_r / \tau_{TF}$.  The results of a linear regression are
$a = 0.194\pm0.007$ and $x_0 = 1.39\pm 0.06$. 
In comparison, $x=a_{TF}=0.88$ is the position of peak radial density for
the TF atom.
An overall model of deviation from the TF limit can thus be extracted:
\beq
     v_r(r) = v_r^{GEA}(r) + 0.194 \frac{\tau_{TF}}{\rho} \log{(Z^{1/3}r/1.39)}
\eeq
A similar fit performed
for $v_p$ yields results that agree within the fit standard deviation.  
(Details of the fit of the anomalies in $v_p$ and $v_r$ can be found 
in the Supplemental Material).

	
\begin{figure}[!htbp]\centering
		\includegraphics[width=1.0\linewidth,height=0.41\textheight,keepaspectratio]{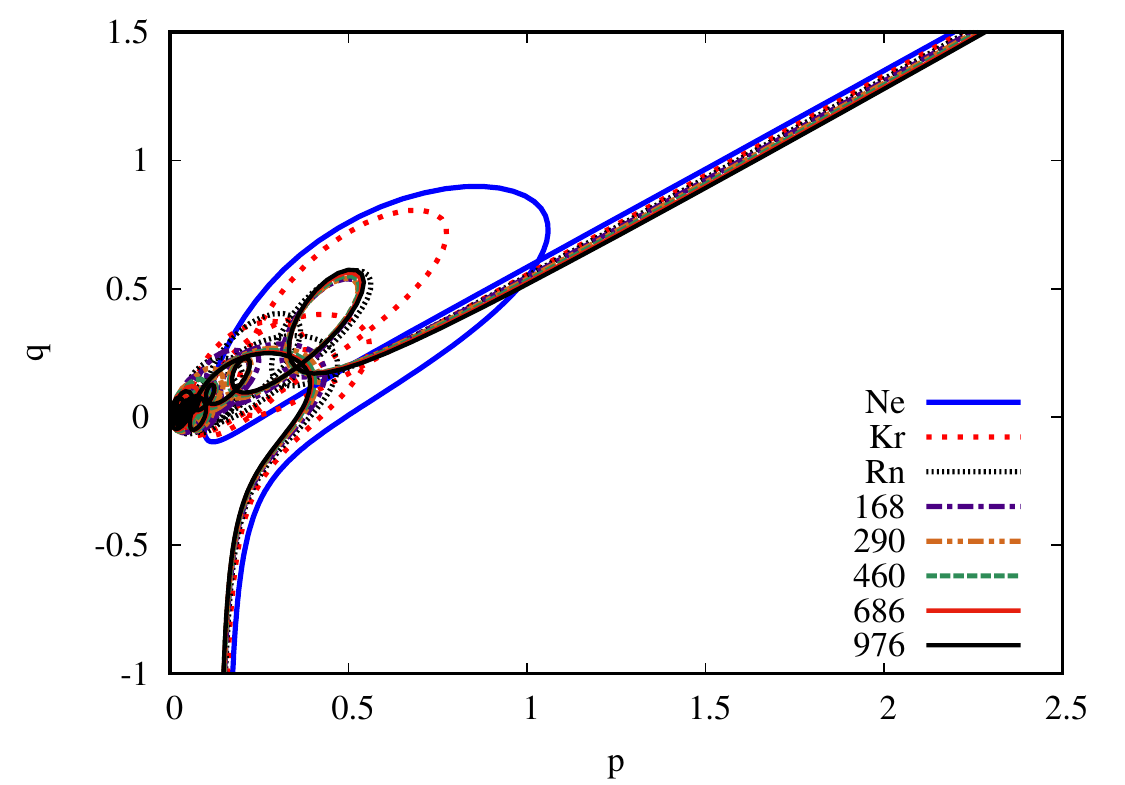}
		\caption{\label{fig:p_vs_q_even}
Parametric plot of $q(r)$ versus $p(r)$ for noble gas atoms with
even principle quantum numbers.}
	\end{figure}

It is worthwhile to analyze this deviation in terms of the
scaled gradient and Laplacian of density in this region.  
One would  
expect these to become small nearly everywhere as $Z\to\infty$ and 
the total energy tends to the TF limit, 
but our 
results with the potential call this expectation into question.
To this end, Fig.~\ref{fig:p_vs_q_even} shows parametric plots of the
scale-invariant quantities 
$p(\bld{r})$ vs $q(\bld{r})$ for noble gases with even principle quantum 
numbers.  Parametric plots for other large-$Z$ atoms can
be found in the Supplemental Material.

There are three clear regions of behavior in this 
plot.\cite{thesis, cancioredd}
The asymptotic approach of $q(\bld{r})\to -\infty$ indicates the 
nuclear cusp. 
The tail where $p(\bld{r})$ and $q(\bld{r})$ both tend to infinity
indicates the evanescent region far from the atom. 
The loops or ``orbits'' come from the atomic core, 
reflecting the oscillations in $F_p$ and $v_p$ seen in Figs.~
\ref{fig:Fresp_Rn} and~\ref{fig:Fresp_976}. 
Each orbit represents a new shell, with $p$ and $q$
moving outwards to their largest values in regions between shells,
and approaching the TF limit $p\!=q\!=0$ in the center of each shell.  
Thus one loop is seen for Ne with two shells.
For increasing $Z$ one may
in general see most loops shrinking towards $p\!=q\!=0$, 
indicating that the TF limit is being approached locally. 
But surprisingly, the process 
stops for the outermost loop, starting with radon.  
Subsequently this outer loop, formed by the valence shell and transition 
to the next shell is largely invariant with atomic number.  A second loop 
seems largely stabilized by $Z=168$, and so on.
For element 976, the inner shells are all close to the TF limit.  
However the outer orbitals gradually deviate from the TF limit, 
and show no sign of ever converging to this limit as $Z \to \infty$.
So the deviation from the TF limit of $v_r$ for $Z^{1/3}>1$ is
indicated by a similar deviation in the GE variables, suggesting
that the GE will never become accurate for these shells.
Similar behavior can be seen for other columns of the periodic table, with
considerable differences in the last shell or two -- data for columns
2, 12 and 13 of the periodic table are shown in the Supplemental Material.

		
It is interesting to note that Eq~\ref{eq:tsqrtrho} implies that
\beq
	\label{eq:vpts}
	v_p=\mu-v_{KS}-v_{vW},
\eeq
where $v_{vW} = \frac{\delta T_{vW}}{\delta\rho}$ and $\mu\sim 0$ may
be taken for a large-$Z$ atom.  This implies that the 
unexpected behavior in the Pauli potential ought to mirrored in the KS
potential as well.  We find that this is indeed the case, calculating
the XC potential using  
the Leeuwen-Baerends exchange potential~\cite{van1994exchange}, 
and the Perdew-Zunger LDA correlation potential~\cite{PerdewZunger}.
The combined Hartree plus XC potential veers off the 
TF limit of $-5\tau_{TF}/3\rho$, to nearly cancel the Pauli potential, with each piece contributing about half of the net effect. The vW contribution 
does not show any unexpected behavior.

\subsection{Evanescent Region}

\begin{figure}[!htbp]\centering
		\includegraphics[width=1.0\linewidth,height=0.41\textheight,keepaspectratio]{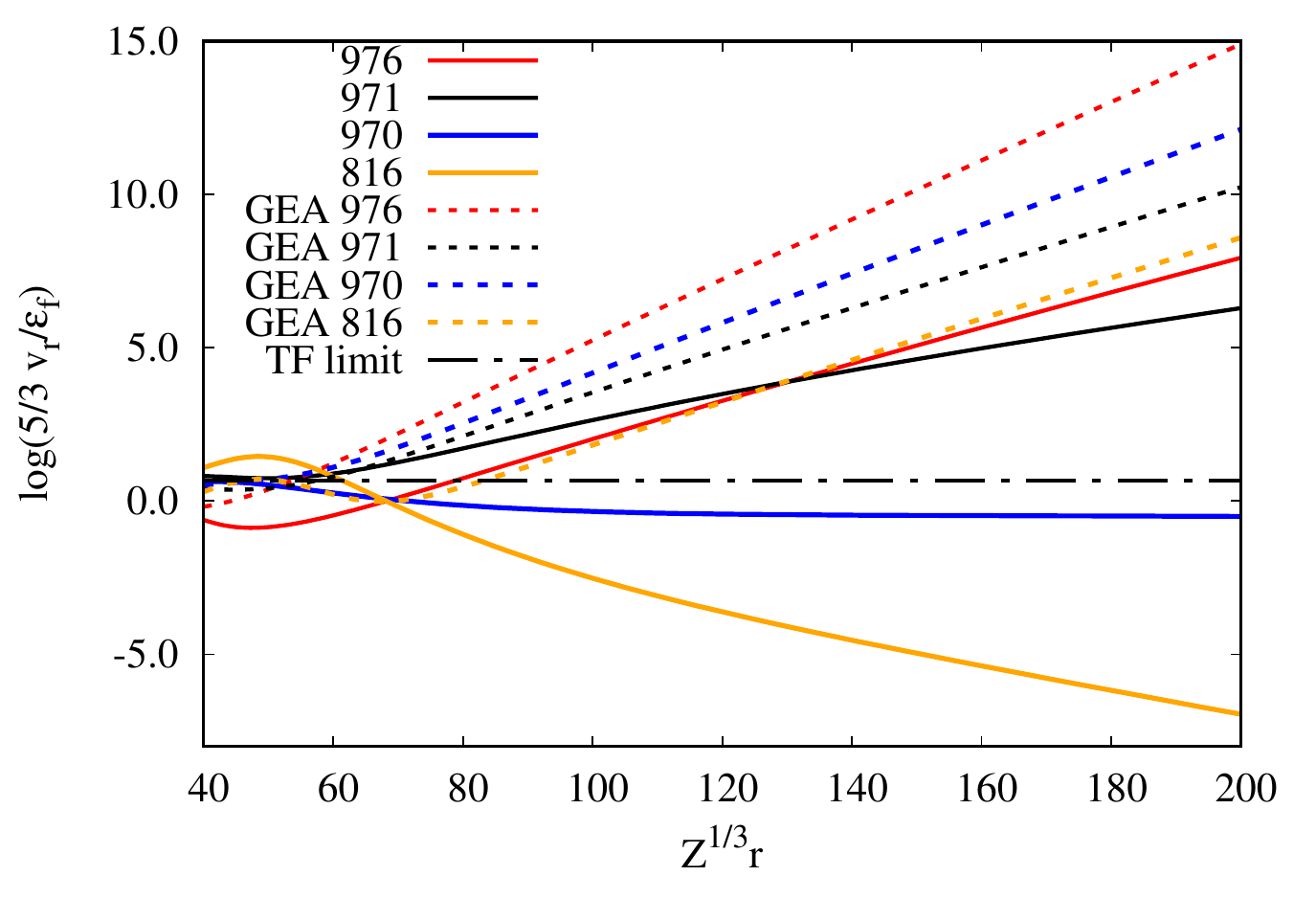}
		\caption{\label{fig:tail_row_16}Evanescent behavior of $v_r$ for the alkali atom ($\!=\!816$), closed d shell atom (970), column 13 atom (971), and noble gas atom (976) with highest principle quantum number 16, as well as 
the GEA for each (dotted lines).}

	\end{figure}

Fig.~\ref{fig:tail_row_16} allows one to examine the evanescent region 
more closely. 
It plots the response potentials for elements 
976 (red), 971 (black), 970 (blue), and 816 (gold), that is,
the atom for column 2, 12, 13 and 18 for the 16th row of the 
extended nonrelativistic periodic table. 
Elements 976 and 971 both have p shells as their highest occupied atomic 
orbital (HOAO) and tend asymptotically to infinity.  
Elements 970 and 816 both have s shells as their HOAO and tend to zero.  
(Thus elements with s shells as their HOAO lack the last local minimum
in $v_r$ present in the other cases.) 
The same trends are shown by the Pauli potential and the Pauli 
enhancement factor $F_p$.

The standard GEA to the response potential for each element is 
shown with the same color and a dashed line.  
Other GE-like models (Loc, Airy gas) make very similar predictions.
Note that the GEA shows roughly the
same limiting behavior regardless of the column of the periodic table. 
They make reasonably accurate predictions for elements that have 
p shells as their HOAO but completely fail for the others.  
 
One can partly explain the asymptotic trend of $v_r$ as $r\to \infty$ 
with a simple analysis of Eq.~\ref{eq:vrespexact}.  The highest energy
shell that contributes to an atom with $M$ shells is the $M-1$ shell.  
Defining $\rho_M$ as the density of the $M$-th energy shell,
the assumption that $\rho(r)\approx\rho_M(r)$ as $r\to\infty$ 
gives the following approximation:
	\beq
		\label{eq:vrevanend}
		\lim_{r\to\infty} \frac{\rho(r) v_r(r)}{\tau_{TF}(r)}
\approx\frac{2(\epsilon_M-\epsilon_{M-1})\rho_{M-1}(r)}{[\rho_M(r)]^{5/3}}.
	\eeq
The slope of the asymptotic behavior 
of Fig.~\ref{fig:tail_row_16} should thus be predicted by the logarithm
of this result.
The long-range exponential decay constant for the HOAO orbital is proportional 
to the square root of its eigenvalue, $\sqrt{2|\epsilon_M|}$, and this 
dominates the behavior of the density in the evanescent region.~\cite{mparrlevy}
A reasonable expectation for the decay constant for the second HOAO, at least for
the local or semilocal exchange models employed in most DFT's is, 
$\sqrt{2|\epsilon_{M-1}|}$.~\cite{evanescent}  In this case, the roughly 
linear behavior in Fig.~\ref{fig:tail_row_16} can be explained by a decay rate 
$\kappa \sim \sqrt{|\epsilon_{M-1}|} - (5/3)\sqrt{|\epsilon_{M}|}$,
as shown in Table~\ref{table:evanescenteigen}.  The 
predicted rates agree closely with the observed behavior in 
Fig.~\ref{fig:tail_row_16}.  In 
general, the main predictor of the evanescent rate is whether the
HOAO and second HOAO have the same or different principal quantum numbers.
The former case leads to a $v_r$ that dies off slowly and tends to 
infinity relative to the local fermi energy $\sim \tau_{TF}/\rho$,
while the latter case shows the opposite effect.
Finally we note that the same qualitative trend in asymptotic behavior
occurs for the full Pauli potential and the additional
term $\tau_P/\rho$ that contributes to it, despite centrifugal terms~\cite{constantin2015atom} that contribute to these quantities. 

 \begin{table}[htb]
 	
 	\caption{\label{table:evanescenteigen} The two highest 
eigenvalues ($\epsilon_M$ and $\epsilon_{M-1}$) belonging to the 
HOAO and second HOAO states respectively, 
and the predicted exponential decay rate $\kappa$, 
for atoms in the 
16th row of the extended periodic table.  Negative $\kappa$ indicates
exponential growth.}
 	\begin{ruledtabular}
 		\begin{tabular}{l|ccc}
 			Atom & $\epsilon_M$ & $\epsilon_{M-1}$ & $\kappa$ \\ \hline\hline
 			816  &-0.0859& -0.4004 & 0.1443\\ \hline
 			970  &-0.1389 & -0.4110 & 0.0199\\ \hline
 			971  &-0.0928 & -0.1827 & -0.0802\\ \hline
 			976  &-0.2113 & -0.3572 & -0.1685\\ 
 		\end{tabular}
 	\end{ruledtabular}
 \end{table}

	\section{Discussion and Conclusions}\label{Conclusions}
In this paper, we have mapped the behavior of Pauli potentials of 
closed-shell atoms up to $Z=976$ or 16 complete energy shells. 
This represents a partial traversal of Lieb-Simon scaling 
to infinite $Z$, a process that transforms the Hamiltonian and expectations 
of real atoms to a limit where Thomas-Fermi theory is relatively exact for
energies. 
Unlike energy expectations, expectations that are functions of position -- 
the electron density and, in this paper, the Pauli potential -- do not have 
to go to the Thomas-Fermi limit uniformly, and the electron density is 
richly structured even in the Thomas-Fermi limit.  
We find that this is true of the Pauli potential as well.

In comparison to the six regions of the large-$Z$ 
limit of electron density defined by Refs.~\onlinecite{liebsimon} and
~\onlinecite{heilmannlieb}, we can identify in our results perhaps five:
\benum
\item There is a near-nuclear region of constant Pauli potential.
	
\item An inner core region consisting of half of the occupied energy 
shells where the potential oscillates about the TF limit.
	
\item An outer core region, where the potential experiences an unexpected
departure from the TF limit.

\item A small valence region where the last oscillation occurs,
where the potential deviates slightly from the anomalous trend, varying
somewhat between columns of the periodic table.
	
\item An evanescent region that 
also varies for each column of the periodic table.  
The slope of this evanescent region is related to the eigenvalues of the 
last two shells.

\eenum

Notably, 
the $1/r^6$ limiting behavior of the TF atom density is just barely hinted at
for the largest atom we study.  
The Lieb-Simon 
limit is understandably harder to reach for local features like the Pauli 
potential than for globally integrated quantities like the kinetic energy.

In the near-nuclear region (1) we find a 
constraint on the Pauli potential, analogous to 
that on the Pauli KED found in Ref.~\onlinecite{constantin2015atom}.
The Pauli potential in the limit $r\to 0$ tends to the difference between the 
highest and lowest occupied eigenvalues.  
This result is consistent with the interpretation of the OFDFT Euler 
equation [Eq.~(\ref{eq:tsqrtrho})] as solving for the density of a system of 
fictitious bosons 
constrained to have the same density as the true fermionic system.
Then the action of the Pauli potential in this region is to shift 
the energy of ones fictitious bosonic system from      
the lowest energy level of the actual fermionic system to that of the 
chemical potential $\mu$.
It is interesting that the response potential in this region is nearly
constant.  Using the theorems of Ref.~\onlinecite{heilmannlieb}, 
it should be possible to prove that the slope of the response potential
and thus the Pauli potential is zero at the nucleus.

The inner core shells are the only region where the Pauli potential
clearly tends to the Thomas-Fermi limit as $Z\to\infty$.  It is 
well fit by the gradient expansion in whatever variant, with the 
best candidate being the Loc GEA, which was fit to the KED in this region. 
This helps justify this model 
because KED by itself is ambiguously defined -- it is the total energy
and the potential that are the physical measurables of the system.  
The key point for an optimal fit seems to be a negative value to the 
coefficient for the $s^2$ term in the expansion --
the Airy gas model with a similar negative coefficient 
performs about as well.  
Nevertheless, the fact that \textit{all} variants of the gradient 
expansion work nearly equally well tells us that the dominant contribution
of the Pauli potential comes from the removal of the von Weizs\"{a}cker
KE from the Kohn-Sham kinetic energy.  In fact, having no gradient 
correction at all, i.e., a Thomas-Fermi energy, would produce quite a
good Pauli potential for atoms, if the real density could be used.
That is, the Thomas-Fermi energy is not so much the problem here
as the Thomas-Fermi density. 

For outer half of the core, there is a surprising deviation from the TF
limit as the system is scaled to large $Z$.  
This effect grows with $Z$ in a consistent way
even as the size of quantum oscillations decreases, 
indicating that the potential in the Lieb-Simon limit diverges from the 
TF potential. 
This process seems not to be an artifact of numerical methods, and 
since it involves half of the energy shells, cannot depend much 
on minor errors in the Aufbau principle for determining the 
order of occupying the outermost shells.  

The fact that the trend occurs over nearly half of the shells 
suggests a clue as to the origin of this effect.
Within the Aufbau principle, the inner core is composed of energy
shells that are completely filled; that is, all possible angular
momentum subshells of a given quantum number are filled. 
The outer core shells are incrementally less complete, and the shells 
become dominated by orbitals with large numbers of radial nodes. 
The system structurally slowly trends from a fully three-dimensional, 
homogeneous gas, and towards something like the radial variant of a 
one-dimensional gas.
In fact, if we associate each oscillation in the Pauli potential
with a separate energy shell, the observed discrepancy is linear in the 
number of unoccupied angular momentum subshells for that energy.  This is
zero for the first roughly half of the energy shells of a noble gas atom,
and increases linearly with each shell beyond that -- exactly the behavior
observed here.

This deviation also raises an interesting issue -- does the Lieb-Simon limit 
for the total energy have a local equivalent for the potential, 
and if so where?  For the large-$Z$ atom, it is common knowledge that 
the density diverges from the TF limit at
the nucleus and asymptotically.  Our work indicates that there is
 a finite but not ignorable deviation of the Pauli potential 
for a finite fraction of electrons.  

	
The final, evanescent region is poorly described by all GE approximations.  
The response function for normal DFT models has a dependence upon the 
difference between the two highest energy eigenvalues that leads to an 
extreme range of asymptotic behaviors for the Pauli potential. 
The GE, is, at best, roughly comparable to the behavior
of atoms with a $p$-shell valence and fails to capture the 
asymptotic trend of any other system.



There are a number of potential avenues along which to take this work 
further.  
One obvious track is to model the various deviations of the Pauli
potential from the GE prediction in large-$Z$ atoms. 
A good question would be how to 
implement the near-nucleus constraint defined in Sec.~\ref{sec:nearnucleus}. 
This does not seem possible in a standard semilocal or one-point model
for the Pauli kinetic energy, at least using only the local density,
gradient and Laplacian.
It may be possible however to construct an accurate 
correction that explicitly depends on the nuclear charge $Z$ in the spirit
of Ref.~\onlinecite{acharya}.
Such a correction could not be 
easily made self-consistent, but might be worth the loss of self-consistency
to produce a physically reasonable potential in this region.

Secondly, deriving a GGA model for the
deviation of the Pauli potential from the TF limit in the outer core 
would be of interest, since this region is more directly involved in bonding. 
Preliminary calculations indicate that including fourth-order gradient 
expansion terms or simple generalized gradient approximations fail to 
reproduce the anomalous trend in the potential found for the outer half 
of the core shells. 
Such models can improve the potential in the outermost shell of the atom, 
which may be useful for improving binding, but do not capture
the physics of the atom as a whole.   
The most intractable region to model seems to be the evanescent region, since
the asymptotic behavior of the response potential depends on the eigenvalue 
spectrum.  
But this sensitivity is partly an artifact of the 
character of DFT orbitals -- the asymptotic behavior of 
HF and higher-rung DFT orbitals depends upon the HOMO eigenvalue only, which
may simplify the task for OFDFT considerably.  

Finally, it would be of considerable interest to extend this study to 
relativistic systems.~\cite{Oganesson}  
We have studied nonrelativistic atoms because of the 
well-known large-$Z$ limiting behavior for this case.
For relativistic systems, spin-orbit coupling spreads out eigenvalues
with the same principal quantum number and grows more important 
as $Z$ increases. 
For the largest-$Z$ atoms 
fabricated in the lab, such as Oganesson, it is believed that this effect
kills shell structure altogether in the outer core.~\cite{Jerabek2018} 
As this leads to a more homogeneous density, one might expect 
less deviation of the Pauli potential from the the Thomas-Fermi limit 
in this region than what we report here.
In contrast, we 
expect that the basic physics underlying the value
of the Pauli potential near the nucleus would be unaltered by 
relativistic corrections. The Pauli potential should still be given 
by the difference in energy between highest and lowest occupied eigenvalues,
although of course these would be very different in value from the 
nonrelativistic case.

		
	

\section*{Supplementary Material}
See supplementary material for tables and plots of kinetic energies using
various models discussed in the paper;
fits for the outer core of Pauli and response potentials and additional $p$ versus $q$ 
parametric plots.  Additional data is available on request.

\begin{acknowledgments}
\end{acknowledgments}
The author would like to thank Sam Trickey and Kieron Burke for helpful 
discussions, Eberhard Engel for use of his atomic DFT code, OPMKS, and
Thomas Baker for help with the intricacies of gnuplot.



\begin{thebibliography}{75}%
	\makeatletter
	\providecommand \@ifxundefined [1]{%
		\@ifx{#1\undefined}
	}%
	\providecommand \@ifnum [1]{%
		\ifnum #1\expandafter \@firstoftwo
		\else \expandafter \@secondoftwo
		\fi
	}%
	\providecommand \@ifx [1]{%
		\ifx #1\expandafter \@firstoftwo
		\else \expandafter \@secondoftwo
		\fi
	}%
	\providecommand \natexlab [1]{#1}%
	\providecommand \enquote  [1]{``#1''}%
	\providecommand \bibnamefont  [1]{#1}%
	\providecommand \bibfnamefont [1]{#1}%
	\providecommand \citenamefont [1]{#1}%
	\providecommand \href@noop [0]{\@secondoftwo}%
	\providecommand \href [0]{\begingroup \@sanitize@url \@href}%
	\providecommand \@href[1]{\@@startlink{#1}\@@href}%
	\providecommand \@@href[1]{\endgroup#1\@@endlink}%
	\providecommand \@sanitize@url [0]{\catcode `\\12\catcode `\$12\catcode
		`\&12\catcode `\#12\catcode `\^12\catcode `\_12\catcode `\%12\relax}%
	\providecommand \@@startlink[1]{}%
	\providecommand \@@endlink[0]{}%
	\providecommand \url  [0]{\begingroup\@sanitize@url \@url }%
	\providecommand \@url [1]{\endgroup\@href {#1}{\urlprefix }}%
	\providecommand \urlprefix  [0]{URL }%
	\providecommand \Eprint [0]{\href }%
	\providecommand \doibase [0]{http://dx.doi.org/}%
	\providecommand \selectlanguage [0]{\@gobble}%
	\providecommand \bibinfo  [0]{\@secondoftwo}%
	\providecommand \bibfield  [0]{\@secondoftwo}%
	\providecommand \translation [1]{[#1]}%
	\providecommand \BibitemOpen [0]{}%
	\providecommand \bibitemStop [0]{}%
	\providecommand \bibitemNoStop [0]{.\EOS\space}%
	\providecommand \EOS [0]{\spacefactor3000\relax}%
	\providecommand \BibitemShut  [1]{\csname bibitem#1\endcsname}%
	\let\auto@bib@innerbib\@empty
	\bibitem [{\citenamefont {Kohn}\ and\ \citenamefont {Sham}(1965)}]{KS}%
	\BibitemOpen
	\bibfield  {author} {\bibinfo {author} {\bibfnamefont {W.}~\bibnamefont
			{Kohn}}\ and\ \bibinfo {author} {\bibfnamefont {L.~J.}\ \bibnamefont
			{Sham}},\ }\href {\doibase 10.1103/PhysRev.140.A1133} {\bibfield  {journal}
		{\bibinfo  {journal} {Phys. Rev.}\ }\textbf {\bibinfo {volume} {140}},\
		\bibinfo {pages} {A1133} (\bibinfo {year} {1965})}\BibitemShut {NoStop}%
	\bibitem [{\citenamefont {Martin}(2004)}]{Martin}%
	\BibitemOpen
	\bibfield  {author} {\bibinfo {author} {\bibfnamefont {R.~M.}\ \bibnamefont
			{Martin}},\ }\href {\doibase 10.1017/CBO9780511805769} {\emph {\bibinfo
			{title} {Electronic Structure: Basic Theory and Practical Methods}}}\
	(\bibinfo  {publisher} {Cambridge University Press},\ \bibinfo {year}
	{2004})\BibitemShut {NoStop}%
	\bibitem [{\citenamefont {Pribram-Jones}, \citenamefont {Gross},\ and\
		\citenamefont {Burke}(2015)}]{IRG}%
	\BibitemOpen
	\bibfield  {author} {\bibinfo {author} {\bibfnamefont {A.}~\bibnamefont
			{Pribram-Jones}}, \bibinfo {author} {\bibfnamefont {D.~A.}\ \bibnamefont
			{Gross}}, \ and\ \bibinfo {author} {\bibfnamefont {K.}~\bibnamefont
			{Burke}},\ }\href {\doibase 10.1146/annurev-physchem-040214-121420}
	{\bibfield  {journal} {\bibinfo  {journal} {Annual Review of Physical
				Chemistry}\ }\textbf {\bibinfo {volume} {66}},\ \bibinfo {pages} {283}
		(\bibinfo {year} {2015})}\BibitemShut {NoStop}%
	\bibitem [{\citenamefont {Karasiev}, \citenamefont {Chakraborty},\ and\
		\citenamefont {Trickey}(2013)}]{karasiev}%
	\BibitemOpen
	\bibfield  {author} {\bibinfo {author} {\bibfnamefont {V.}~\bibnamefont
			{Karasiev}}, \bibinfo {author} {\bibfnamefont {D.}~\bibnamefont
			{Chakraborty}}, \ and\ \bibinfo {author} {\bibfnamefont {S.}~\bibnamefont
			{Trickey}},\ }\href {\doibase 10.1007/978-3-319-06379-9_6} {\enquote
		{\bibinfo {title} {Progress on new approaches to old ideas: Orbital-free
				density functionals},}\ } (\bibinfo {year} {2013})\BibitemShut {NoStop}%
	\bibitem [{\citenamefont {Akimov}\ and\ \citenamefont
		{Prezhdo}(2015)}]{akimov2015large}%
	\BibitemOpen
	\bibfield  {author} {\bibinfo {author} {\bibfnamefont {A.~V.}\ \bibnamefont
			{Akimov}}\ and\ \bibinfo {author} {\bibfnamefont {O.~V.}\ \bibnamefont
			{Prezhdo}},\ }\href {\doibase 10.1021/cr500524c} {\bibfield  {journal}
		{\bibinfo  {journal} {Chemical Reviews}\ }\textbf {\bibinfo {volume} {115}},\
		\bibinfo {pages} {5797} (\bibinfo {year} {2015})}\BibitemShut {NoStop}%
	\bibitem [{\citenamefont {Graziani}\ \emph {et~al.}(2014)\citenamefont
		{Graziani}, \citenamefont {Desjarlais}, \citenamefont {Redmer},\ and\
		\citenamefont {Trickey}}]{graziani2014frontiers}%
	\BibitemOpen
	\bibinfo {editor} {\bibfnamefont {F.}~\bibnamefont {Graziani}}, \bibinfo
	{editor} {\bibfnamefont {M.~P.}\ \bibnamefont {Desjarlais}}, \bibinfo
	{editor} {\bibfnamefont {R.}~\bibnamefont {Redmer}}, \ and\ \bibinfo {editor}
	{\bibfnamefont {S.~B.}\ \bibnamefont {Trickey}},\ eds.,\ \href {\doibase
		10.1007/978-3-319-04912-0} {\emph {\bibinfo {title} {Frontiers and Challenges
				in Warm Dense Matter}}},\ \bibinfo {series} {Lecture Notes in Computational
		Science and Engineering}, Vol.~\bibinfo {volume} {96}\ (\bibinfo  {publisher}
	{Springer International Publishing},\ \bibinfo {address} {Switzerland},\
	\bibinfo {year} {2014})\BibitemShut {NoStop}%
	\bibitem [{\citenamefont {Rosner}, \citenamefont {Hammer},\ and\ \citenamefont
		{Rothman}(2010)}]{rosnerbasic}%
	\BibitemOpen
	\bibfield  {author} {\bibinfo {author} {\bibfnamefont {R.}~\bibnamefont
			{Rosner}}, \bibinfo {author} {\bibfnamefont {D.}~\bibnamefont {Hammer}}, \
		and\ \bibinfo {author} {\bibfnamefont {T.}~\bibnamefont {Rothman}},\
	}\href@noop {} {\enquote {\bibinfo {title} {Basic research needs for high
				energy density laboratory physics, report on the workshop on high energy
				density laboratory physics research needs, nov. 15–18, 2009},}\ }\bibinfo
	{type} {Tech. Rep.}\ (\bibinfo  {institution} {US Department of Energy},\
	\bibinfo {address} {Washington, DC},\ \bibinfo {year} {2010})\BibitemShut
	{NoStop}%
	\bibitem [{\citenamefont {Karasiev}, \citenamefont {Sjostrom},\ and\
		\citenamefont {Trickey}(2012)}]{karasievWDM}%
	\BibitemOpen
	\bibfield  {author} {\bibinfo {author} {\bibfnamefont {V.~V.}\ \bibnamefont
			{Karasiev}}, \bibinfo {author} {\bibfnamefont {T.}~\bibnamefont {Sjostrom}},
		\ and\ \bibinfo {author} {\bibfnamefont {S.~B.}\ \bibnamefont {Trickey}},\
	}\href {\doibase 10.1103/PhysRevB.86.115101} {\bibfield  {journal} {\bibinfo
			{journal} {Phys. Rev. B}\ }\textbf {\bibinfo {volume} {86}},\ \bibinfo
		{pages} {115101} (\bibinfo {year} {2012})}\BibitemShut {NoStop}%
	\bibitem [{\citenamefont {Hohenberg}\ and\ \citenamefont {Kohn}(1964)}]{HK}%
	\BibitemOpen
	\bibfield  {author} {\bibinfo {author} {\bibfnamefont {P.}~\bibnamefont
			{Hohenberg}}\ and\ \bibinfo {author} {\bibfnamefont {W.}~\bibnamefont
			{Kohn}},\ }\href {\doibase 10.1103/PhysRev.136.B864} {\bibfield  {journal}
		{\bibinfo  {journal} {Phys. Rev.}\ }\textbf {\bibinfo {volume} {136}},\
		\bibinfo {pages} {B864} (\bibinfo {year} {1964})}\BibitemShut {NoStop}%
	\bibitem [{\citenamefont {Tran}\ and\ \citenamefont
		{Wesolowski}(2013)}]{TranWes}%
	\BibitemOpen
	\bibfield  {author} {\bibinfo {author} {\bibfnamefont {F.}~\bibnamefont
			{Tran}}\ and\ \bibinfo {author} {\bibfnamefont {T.~A.}\ \bibnamefont
			{Wesolowski}}\ }(\bibinfo {year} {2013})\BibitemShut {NoStop}%
	\bibitem [{\citenamefont {Witt}\ \emph {et~al.}(2018)\citenamefont {Witt},
		\citenamefont {del Rio}, \citenamefont {Dieterich},\ and\ \citenamefont
		{Carter}}]{witt2018orbital}%
	\BibitemOpen
	\bibfield  {author} {\bibinfo {author} {\bibfnamefont {W.~C.}\ \bibnamefont
			{Witt}}, \bibinfo {author} {\bibfnamefont {B.~G.}\ \bibnamefont {del Rio}},
		\bibinfo {author} {\bibfnamefont {J.~M.}\ \bibnamefont {Dieterich}}, \ and\
		\bibinfo {author} {\bibfnamefont {E.~A.}\ \bibnamefont {Carter}},\ }\href
	{\doibase 10.1557/jmr.2017.462} {\bibfield  {journal} {\bibinfo  {journal}
			{Journal of Materials Research}\ }\textbf {\bibinfo {volume} {33}},\ \bibinfo
		{pages} {777} (\bibinfo {year} {2018})}\BibitemShut {NoStop}%
	\bibitem [{\citenamefont {Mejia-Rodriguez}\ and\ \citenamefont
		{Trickey}(2018)}]{trickey_deorbitalization}%
	\BibitemOpen
	\bibfield  {author} {\bibinfo {author} {\bibfnamefont {D.}~\bibnamefont
			{Mejia-Rodriguez}}\ and\ \bibinfo {author} {\bibfnamefont {S.~B.}\
			\bibnamefont {Trickey}},\ }\href {\doibase 10.1103/PhysRevB.98.115161}
	{\bibfield  {journal} {\bibinfo  {journal} {Phys. Rev. B}\ }\textbf {\bibinfo
			{volume} {98}},\ \bibinfo {pages} {115161} (\bibinfo {year}
		{2018})}\BibitemShut {NoStop}%
	\bibitem [{\citenamefont {{T}homas}(1927)}]{thomas1927calculation}%
	\BibitemOpen
	\bibfield  {author} {\bibinfo {author} {\bibfnamefont {L.~H.}\ \bibnamefont
			{{T}homas}},\ }\href {\doibase 10.1017/S0305004100011683} {\bibfield
		{journal} {\bibinfo  {journal} {Mathematical Proceedings of the Cambridge
				Philosophical Society}\ }\textbf {\bibinfo {volume} {23}},\ \bibinfo {pages}
		{542} (\bibinfo {year} {1927})}\BibitemShut {NoStop}%
	\bibitem [{\citenamefont {Fermi}(1927)}]{F27}%
	\BibitemOpen
	\bibfield  {author} {\bibinfo {author} {\bibfnamefont {E.}~\bibnamefont
			{Fermi}},\ }\href@noop {} {\bibfield  {journal} {\bibinfo  {journal} {Rend.
				Acc. Naz. Lincei}\ }\textbf {\bibinfo {volume} {6}} (\bibinfo {year}
		{1927})}\BibitemShut {NoStop}%
	\bibitem [{\citenamefont {Teller}(1962)}]{teller}%
	\BibitemOpen
	\bibfield  {author} {\bibinfo {author} {\bibfnamefont {E.}~\bibnamefont
			{Teller}},\ }\href {\doibase 10.1103/RevModPhys.34.627} {\bibfield  {journal}
		{\bibinfo  {journal} {Rev. Mod. Phys.}\ }\textbf {\bibinfo {volume} {34}},\
		\bibinfo {pages} {627} (\bibinfo {year} {1962})}\BibitemShut {NoStop}%
	\bibitem [{\citenamefont {Xia}\ \emph {et~al.}(2012)\citenamefont {Xia},
		\citenamefont {Huang}, \citenamefont {Shin},\ and\ \citenamefont
		{Carter}}]{cartermol}%
	\BibitemOpen
	\bibfield  {author} {\bibinfo {author} {\bibfnamefont {J.}~\bibnamefont
			{Xia}}, \bibinfo {author} {\bibfnamefont {C.}~\bibnamefont {Huang}}, \bibinfo
		{author} {\bibfnamefont {I.}~\bibnamefont {Shin}}, \ and\ \bibinfo {author}
		{\bibfnamefont {E.~A.}\ \bibnamefont {Carter}},\ }\href {\doibase
		10.1063/1.3685604} {\bibfield  {journal} {\bibinfo  {journal} {The Journal of
				Chemical Physics}\ }\textbf {\bibinfo {volume} {136}},\ \bibinfo {pages}
		{084102} (\bibinfo {year} {2012})}\BibitemShut {NoStop}%
	\bibitem [{\citenamefont {Finzel}(2018)}]{finzel2018chemical}%
	\BibitemOpen
	\bibfield  {author} {\bibinfo {author} {\bibfnamefont {K.}~\bibnamefont
			{Finzel}},\ }\href {\doibase https://doi.org/10.1016/j.comptc.2018.10.004}
	{\bibfield  {journal} {\bibinfo  {journal} {Computational and Theoretical
				Chemistry}\ }\textbf {\bibinfo {volume} {1144}},\ \bibinfo {pages} {50 }
		(\bibinfo {year} {2018})}\BibitemShut {NoStop}%
	\bibitem [{\citenamefont {Tran}\ and\ \citenamefont
		{Wesolowski}(2002)}]{TranWesolowski}%
	\BibitemOpen
	\bibfield  {author} {\bibinfo {author} {\bibfnamefont {F.}~\bibnamefont
			{Tran}}\ and\ \bibinfo {author} {\bibfnamefont {T.~A.}\ \bibnamefont
			{Wesolowski}},\ }\href {\doibase 10.1002/qua.10306} {\bibfield  {journal}
		{\bibinfo  {journal} {Int. J. Quantum Chem.}\ }\textbf {\bibinfo {volume}
			{89}},\ \bibinfo {pages} {441} (\bibinfo {year} {2002})}\BibitemShut
	{NoStop}%
	\bibitem [{\citenamefont {Lacks}\ and\ \citenamefont
		{Gordon}(1994)}]{LacksGordon94}%
	\BibitemOpen
	\bibfield  {author} {\bibinfo {author} {\bibfnamefont {D.~J.}\ \bibnamefont
			{Lacks}}\ and\ \bibinfo {author} {\bibfnamefont {R.~G.}\ \bibnamefont
			{Gordon}},\ }\href {\doibase http://dx.doi.org/10.1063/1.466274} {\bibfield
		{journal} {\bibinfo  {journal} {J. Chem. Phys.}\ }\textbf {\bibinfo {volume}
			{100}},\ \bibinfo {pages} {4446} (\bibinfo {year} {1994})}\BibitemShut
	{NoStop}%
	\bibitem [{\citenamefont {Thakkar}(1992)}]{Thakkar92}%
	\BibitemOpen
	\bibfield  {author} {\bibinfo {author} {\bibfnamefont {A.}~\bibnamefont
			{Thakkar}},\ }\href {\doibase 10.1103/PhysRevA.46.6920} {\bibfield  {journal}
		{\bibinfo  {journal} {Phys. Rev. A}\ }\textbf {\bibinfo {volume} {46}},\
		\bibinfo {pages} {6920} (\bibinfo {year} {1992})}\BibitemShut {NoStop}%
	\bibitem [{\citenamefont {Constantin}\ \emph {et~al.}(2011)\citenamefont
		{Constantin}, \citenamefont {Fabiano}, \citenamefont {Laricchia},\ and\
		\citenamefont {Della~Sala}}]{APBE}%
	\BibitemOpen
	\bibfield  {author} {\bibinfo {author} {\bibfnamefont {L.~A.}\ \bibnamefont
			{Constantin}}, \bibinfo {author} {\bibfnamefont {E.}~\bibnamefont {Fabiano}},
		\bibinfo {author} {\bibfnamefont {S.}~\bibnamefont {Laricchia}}, \ and\
		\bibinfo {author} {\bibfnamefont {F.}~\bibnamefont {Della~Sala}},\ }\href
	{\doibase 10.1103/PhysRevLett.106.186406} {\bibfield  {journal} {\bibinfo
			{journal} {Phys. Rev. Lett.}\ }\textbf {\bibinfo {volume} {106}},\ \bibinfo
		{pages} {186406} (\bibinfo {year} {2011})}\BibitemShut {NoStop}%
	\bibitem [{\citenamefont {Karasiev}\ \emph {et~al.}(2013)\citenamefont
		{Karasiev}, \citenamefont {Chakraborty}, \citenamefont {Shukruto},\ and\
		\citenamefont {Trickey}}]{VT84F}%
	\BibitemOpen
	\bibfield  {author} {\bibinfo {author} {\bibfnamefont {V.~V.}\ \bibnamefont
			{Karasiev}}, \bibinfo {author} {\bibfnamefont {D.}~\bibnamefont
			{Chakraborty}}, \bibinfo {author} {\bibfnamefont {O.~A.}\ \bibnamefont
			{Shukruto}}, \ and\ \bibinfo {author} {\bibfnamefont {S.~B.}\ \bibnamefont
			{Trickey}},\ }\href {\doibase 10.1103/PhysRevB.88.161108} {\bibfield
		{journal} {\bibinfo  {journal} {Phys. Rev. B}\ }\textbf {\bibinfo {volume}
			{88}},\ \bibinfo {pages} {161108} (\bibinfo {year} {2013})}\BibitemShut
	{NoStop}%
	\bibitem [{\citenamefont {Borgoo}, \citenamefont {Green},\ and\ \citenamefont
		{Tozer}(2014)}]{BorgooMol}%
	\BibitemOpen
	\bibfield  {author} {\bibinfo {author} {\bibfnamefont {A.}~\bibnamefont
			{Borgoo}}, \bibinfo {author} {\bibfnamefont {J.~A.}\ \bibnamefont {Green}}, \
		and\ \bibinfo {author} {\bibfnamefont {D.~J.}\ \bibnamefont {Tozer}},\ }\href
	{\doibase 10.1021/ct500670h} {\bibfield  {journal} {\bibinfo  {journal}
			{Journal of Chemical Theory and Computation}\ }\textbf {\bibinfo {volume}
			{10}},\ \bibinfo {pages} {5338} (\bibinfo {year} {2014})}\BibitemShut
	{NoStop}%
	\bibitem [{\citenamefont {Luo}, \citenamefont {Karasiev},\ and\ \citenamefont
		{Trickey}(2018)}]{LKT}%
	\BibitemOpen
	\bibfield  {author} {\bibinfo {author} {\bibfnamefont {K.}~\bibnamefont
			{Luo}}, \bibinfo {author} {\bibfnamefont {V.~V.}\ \bibnamefont {Karasiev}}, \
		and\ \bibinfo {author} {\bibfnamefont {S.~B.}\ \bibnamefont {Trickey}},\
	}\href {\doibase 10.1103/PhysRevB.98.041111} {\bibfield  {journal} {\bibinfo
			{journal} {Phys. Rev. B}\ }\textbf {\bibinfo {volume} {98}},\ \bibinfo
		{pages} {041111} (\bibinfo {year} {2018})}\BibitemShut {NoStop}%
	\bibitem [{\citenamefont {Perdew}\ and\ \citenamefont {Constantin}(2007)}]{PC}%
	\BibitemOpen
	\bibfield  {author} {\bibinfo {author} {\bibfnamefont {J.~P.}\ \bibnamefont
			{Perdew}}\ and\ \bibinfo {author} {\bibfnamefont {L.~A.}\ \bibnamefont
			{Constantin}},\ }\href {\doibase 10.1103/PhysRevB.75.155109} {\bibfield
		{journal} {\bibinfo  {journal} {Phys. Rev. B}\ }\textbf {\bibinfo {volume}
			{75}},\ \bibinfo {pages} {155109} (\bibinfo {year} {2007})}\BibitemShut
	{NoStop}%
	\bibitem [{\citenamefont {Lee}\ \emph {et~al.}(2009)\citenamefont {Lee},
		\citenamefont {Constantin}, \citenamefont {Perdew},\ and\ \citenamefont
		{Burke}}]{LCPB}%
	\BibitemOpen
	\bibfield  {author} {\bibinfo {author} {\bibfnamefont {D.}~\bibnamefont
			{Lee}}, \bibinfo {author} {\bibfnamefont {L.~A.}\ \bibnamefont {Constantin}},
		\bibinfo {author} {\bibfnamefont {J.~P.}\ \bibnamefont {Perdew}}, \ and\
		\bibinfo {author} {\bibfnamefont {K.}~\bibnamefont {Burke}},\ }\href
	{\doibase DOI:10.1063/1.3059783} {\bibfield  {journal} {\bibinfo  {journal}
			{J. Chem. Phys.}\ }\textbf {\bibinfo {volume} {130}},\ \bibinfo {pages}
		{034107} (\bibinfo {year} {2009})}\BibitemShut {NoStop}%
	\bibitem [{\citenamefont {Cancio}\ and\ \citenamefont
		{Redd}(2017)}]{cancioredd}%
	\BibitemOpen
	\bibfield  {author} {\bibinfo {author} {\bibfnamefont {A.~C.}\ \bibnamefont
			{Cancio}}\ and\ \bibinfo {author} {\bibfnamefont {J.~J.}\ \bibnamefont
			{Redd}},\ }\href {\doibase 10.1080/00268976.2016.1246757} {\bibfield
		{journal} {\bibinfo  {journal} {Molecular Physics}\ }\textbf {\bibinfo
			{volume} {115}},\ \bibinfo {pages} {618} (\bibinfo {year} {2017})},\ \Eprint
	{http://arxiv.org/abs/https://doi.org/10.1080/00268976.2016.1246757}
	{https://doi.org/10.1080/00268976.2016.1246757} \BibitemShut {NoStop}%
	\bibitem [{\citenamefont {Constantin}, \citenamefont {Fabiano},\ and\
		\citenamefont {Della~Sala}(2018)}]{CFS2018}%
	\BibitemOpen
	\bibfield  {author} {\bibinfo {author} {\bibfnamefont {L.~A.}\ \bibnamefont
			{Constantin}}, \bibinfo {author} {\bibfnamefont {E.}~\bibnamefont {Fabiano}},
		\ and\ \bibinfo {author} {\bibfnamefont {F.}~\bibnamefont {Della~Sala}},\
	}\href {\doibase 10.1021/acs.jpclett.8b01926} {\bibfield  {journal} {\bibinfo
			{journal} {The Journal of Physical Chemistry Letters}\ }\textbf {\bibinfo
			{volume} {9}},\ \bibinfo {pages} {4385} (\bibinfo {year} {2018})}\BibitemShut
	{NoStop}%
	\bibitem [{\citenamefont {Constantin}, \citenamefont {Fabiano},\ and\
		\citenamefont {Della~Sala}(2017{\natexlab{a}})}]{CFS20194thorder}%
	\BibitemOpen
	\bibfield  {author} {\bibinfo {author} {\bibfnamefont {L.~A.}\ \bibnamefont
			{Constantin}}, \bibinfo {author} {\bibfnamefont {E.}~\bibnamefont {Fabiano}},
		\ and\ \bibinfo {author} {\bibfnamefont {F.}~\bibnamefont {Della~Sala}},\
	}\href {\doibase 10.1021/acs.jctc.7b00705} {\bibfield  {journal} {\bibinfo
			{journal} {Journal of Chemical Theory and Computation}\ }\textbf {\bibinfo
			{volume} {13}},\ \bibinfo {pages} {4228} (\bibinfo {year}
		{2017}{\natexlab{a}})}\BibitemShut {NoStop}%
	\bibitem [{\citenamefont {Constantin}, \citenamefont {Fabiano},\ and\
		\citenamefont {Della~Sala}(2017{\natexlab{b}})}]{constantin2017modified}%
	\BibitemOpen
	\bibfield  {author} {\bibinfo {author} {\bibfnamefont {L.~A.}\ \bibnamefont
			{Constantin}}, \bibinfo {author} {\bibfnamefont {E.}~\bibnamefont {Fabiano}},
		\ and\ \bibinfo {author} {\bibfnamefont {F.}~\bibnamefont {Della~Sala}},\
	}\href {\doibase 10.1021/acs.jctc.7b00705} {\bibfield  {journal} {\bibinfo
			{journal} {Journal of Chemical Theory and Computation}\ }\textbf {\bibinfo
			{volume} {13}},\ \bibinfo {pages} {4228} (\bibinfo {year}
		{2017}{\natexlab{b}})}\BibitemShut {NoStop}%
	\bibitem [{\citenamefont {Wang}, \citenamefont {Govind},\ and\ \citenamefont
		{Carter}(1999)}]{WangCarterNew}%
	\BibitemOpen
	\bibfield  {author} {\bibinfo {author} {\bibfnamefont {Y.~A.}\ \bibnamefont
			{Wang}}, \bibinfo {author} {\bibfnamefont {N.}~\bibnamefont {Govind}}, \ and\
		\bibinfo {author} {\bibfnamefont {E.~A.}\ \bibnamefont {Carter}},\ }\href
	{\doibase 10.1103/PhysRevB.60.16350} {\bibfield  {journal} {\bibinfo
			{journal} {Phys. Rev. B}\ }\textbf {\bibinfo {volume} {60}},\ \bibinfo
		{pages} {16350} (\bibinfo {year} {1999})}\BibitemShut {NoStop}%
	\bibitem [{\citenamefont {Huang}\ and\ \citenamefont
		{Carter}(2010)}]{huangcarter}%
	\BibitemOpen
	\bibfield  {author} {\bibinfo {author} {\bibfnamefont {C.}~\bibnamefont
			{Huang}}\ and\ \bibinfo {author} {\bibfnamefont {E.~A.}\ \bibnamefont
			{Carter}},\ }\href {\doibase 10.1103/PhysRevB.81.045206} {\bibfield
		{journal} {\bibinfo  {journal} {Phys. Rev. B}\ }\textbf {\bibinfo {volume}
			{81}},\ \bibinfo {pages} {045206} (\bibinfo {year} {2010})}\BibitemShut
	{NoStop}%
	\bibitem [{\citenamefont {Mi}, \citenamefont {Genova},\ and\ \citenamefont
		{Pavanello}(2018)}]{pavanello}%
	\BibitemOpen
	\bibfield  {author} {\bibinfo {author} {\bibfnamefont {W.}~\bibnamefont
			{Mi}}, \bibinfo {author} {\bibfnamefont {A.}~\bibnamefont {Genova}}, \ and\
		\bibinfo {author} {\bibfnamefont {M.}~\bibnamefont {Pavanello}},\ }\href
	{\doibase 10.1063/1.5023926} {\bibfield  {journal} {\bibinfo  {journal} {The
				Journal of Chemical Physics}\ }\textbf {\bibinfo {volume} {148}},\ \bibinfo
		{pages} {184107} (\bibinfo {year} {2018})}\BibitemShut {NoStop}%
	\bibitem [{\citenamefont {Constantin}(2019)}]{ConstLindhard}%
	\BibitemOpen
	\bibfield  {author} {\bibinfo {author} {\bibfnamefont {L.~A.}\ \bibnamefont
			{Constantin}},\ }\href {\doibase 10.1103/PhysRevB.99.155137} {\bibfield
		{journal} {\bibinfo  {journal} {Phys. Rev. B}\ }\textbf {\bibinfo {volume}
			{99}},\ \bibinfo {pages} {155137} (\bibinfo {year} {2019})}\BibitemShut
	{NoStop}%
	\bibitem [{\citenamefont {Levy}\ and\ \citenamefont
		{Ou-Yang}(1988)}]{levy1988exact}%
	\BibitemOpen
	\bibfield  {author} {\bibinfo {author} {\bibfnamefont {M.}~\bibnamefont
			{Levy}}\ and\ \bibinfo {author} {\bibfnamefont {H.}~\bibnamefont {Ou-Yang}},\
	}\href {\doibase 10.1103/PhysRevA.38.625} {\bibfield  {journal} {\bibinfo
			{journal} {Phys. Rev. A}\ }\textbf {\bibinfo {volume} {38}},\ \bibinfo
		{pages} {625} (\bibinfo {year} {1988})}\BibitemShut {NoStop}%
	\bibitem [{\citenamefont {Gritsenko}, \citenamefont {van Leeuwen},\ and\
		\citenamefont {Baerends}(1994)}]{gritsenko1994}%
	\BibitemOpen
	\bibfield  {author} {\bibinfo {author} {\bibfnamefont {O.}~\bibnamefont
			{Gritsenko}}, \bibinfo {author} {\bibfnamefont {R.}~\bibnamefont {van
				Leeuwen}}, \ and\ \bibinfo {author} {\bibfnamefont {E.~J.}\ \bibnamefont
			{Baerends}},\ }\href {\doibase 10.1063/1.468024} {\bibfield  {journal}
		{\bibinfo  {journal} {The Journal of Chemical Physics}\ }\textbf {\bibinfo
			{volume} {101}},\ \bibinfo {pages} {8955} (\bibinfo {year} {1994})},\ \Eprint
	{http://arxiv.org/abs/https://doi.org/10.1063/1.468024}
	{https://doi.org/10.1063/1.468024} \BibitemShut {NoStop}%
	\bibitem [{\citenamefont {{van Leeuwen}}, \citenamefont {{Gritsenko}},\ and\
		\citenamefont {{Baerends}}(1995)}]{van1995step}%
	\BibitemOpen
	\bibfield  {author} {\bibinfo {author} {\bibfnamefont {R.}~\bibnamefont {{van
					Leeuwen}}}, \bibinfo {author} {\bibfnamefont {O.}~\bibnamefont
			{{Gritsenko}}}, \ and\ \bibinfo {author} {\bibfnamefont {E.~J.}\ \bibnamefont
			{{Baerends}}},\ }\href {\doibase 10.1007/BF01437503} {\bibfield  {journal}
		{\bibinfo  {journal} {Zeitschrift fur Physik D Atoms Molecules Clusters}\
		}\textbf {\bibinfo {volume} {33}},\ \bibinfo {pages} {229} (\bibinfo {year}
		{1995})}\BibitemShut {NoStop}%
	\bibitem [{\citenamefont {Baerends}\ and\ \citenamefont
		{Gritsenko}(1997)}]{baerends1997quantum}%
	\BibitemOpen
	\bibfield  {author} {\bibinfo {author} {\bibfnamefont {E.~J.}\ \bibnamefont
			{Baerends}}\ and\ \bibinfo {author} {\bibfnamefont {O.~V.}\ \bibnamefont
			{Gritsenko}},\ }\href {\doibase 10.1021/jp9703768} {\bibfield  {journal}
		{\bibinfo  {journal} {The Journal of Physical Chemistry A}\ }\textbf
		{\bibinfo {volume} {101}},\ \bibinfo {pages} {5383} (\bibinfo {year}
		{1997})},\ \Eprint {http://arxiv.org/abs/https://doi.org/10.1021/jp9703768}
	{https://doi.org/10.1021/jp9703768} \BibitemShut {NoStop}%
	\bibitem [{\citenamefont {Kraisler}\ and\ \citenamefont
		{Schild}(2020)}]{KraislerAxel20}%
	\BibitemOpen
	\bibfield  {author} {\bibinfo {author} {\bibfnamefont {E.}~\bibnamefont
			{Kraisler}}\ and\ \bibinfo {author} {\bibfnamefont {A.}~\bibnamefont
			{Schild}},\ }\href {\doibase 10.1103/PhysRevResearch.2.013159} {\bibfield
		{journal} {\bibinfo  {journal} {Phys. Rev. Research}\ }\textbf {\bibinfo
			{volume} {2}},\ \bibinfo {pages} {013159} (\bibinfo {year}
		{2020})}\BibitemShut {NoStop}%
	\bibitem [{\citenamefont {Finzel}(2020)}]{FinzelMolecules20}%
	\BibitemOpen
	\bibfield  {author} {\bibinfo {author} {\bibfnamefont {K.}~\bibnamefont
			{Finzel}},\ }\href {\doibase 10.3390/molecules25081771} {\bibfield  {journal}
		{\bibinfo  {journal} {Molecules}\ }\textbf {\bibinfo {volume} {25}} (\bibinfo
		{year} {2020}),\ 10.3390/molecules25081771}\BibitemShut {NoStop}%
	\bibitem [{\citenamefont {Finzel}(2019)}]{Finzel19}%
	\BibitemOpen
	\bibfield  {author} {\bibinfo {author} {\bibfnamefont {K.}~\bibnamefont
			{Finzel}},\ }\href {\doibase 10.1063/1.5099217} {\bibfield  {journal}
		{\bibinfo  {journal} {The Journal of Chemical Physics}\ }\textbf {\bibinfo
			{volume} {151}},\ \bibinfo {pages} {024109} (\bibinfo {year}
		{2019})}\BibitemShut {NoStop}%
	\bibitem [{\citenamefont {Lieb}\ and\ \citenamefont {Simon}(1973)}]{liebsimon}%
	\BibitemOpen
	\bibfield  {author} {\bibinfo {author} {\bibfnamefont {E.~H.}\ \bibnamefont
			{Lieb}}\ and\ \bibinfo {author} {\bibfnamefont {B.}~\bibnamefont {Simon}},\
	}\href {\doibase 10.1103/PhysRevLett.31.681} {\bibfield  {journal} {\bibinfo
			{journal} {Phys. Rev. Lett.}\ }\textbf {\bibinfo {volume} {31}},\ \bibinfo
		{pages} {681} (\bibinfo {year} {1973})}\BibitemShut {NoStop}%
	\bibitem [{\citenamefont {Burke}\ \emph {et~al.}(2016)\citenamefont {Burke},
		\citenamefont {Cancio}, \citenamefont {Gould},\ and\ \citenamefont
		{Pittalis}}]{BCGP16}%
	\BibitemOpen
	\bibfield  {author} {\bibinfo {author} {\bibfnamefont {K.}~\bibnamefont
			{Burke}}, \bibinfo {author} {\bibfnamefont {A.}~\bibnamefont {Cancio}},
		\bibinfo {author} {\bibfnamefont {T.}~\bibnamefont {Gould}}, \ and\ \bibinfo
		{author} {\bibfnamefont {S.}~\bibnamefont {Pittalis}},\ }\href {\doibase
		10.1063/1.4959126} {\bibfield  {journal} {\bibinfo  {journal} {The Journal of
				Chemical Physics}\ }\textbf {\bibinfo {volume} {145}},\ \bibinfo {pages}
		{054112} (\bibinfo {year} {2016})}\BibitemShut {NoStop}%
	\bibitem [{\citenamefont {Levy}\ and\ \citenamefont
		{Perdew}(1985)}]{levyperdew}%
	\BibitemOpen
	\bibfield  {author} {\bibinfo {author} {\bibfnamefont {M.}~\bibnamefont
			{Levy}}\ and\ \bibinfo {author} {\bibfnamefont {J.}~\bibnamefont {Perdew}},\
	}\href {\doibase 10.1103/PhysRevA.32.2010} {\bibfield  {journal} {\bibinfo
			{journal} {Phys. Rev. A}\ }\textbf {\bibinfo {volume} {32}},\ \bibinfo
		{pages} {2010} (\bibinfo {year} {1985})}\BibitemShut {NoStop}%
	\bibitem [{\citenamefont {Lieb}\ and\ \citenamefont
		{Simon}(1977)}]{lieb1977thomas}%
	\BibitemOpen
	\bibfield  {author} {\bibinfo {author} {\bibfnamefont {E.~H.}\ \bibnamefont
			{Lieb}}\ and\ \bibinfo {author} {\bibfnamefont {B.}~\bibnamefont {Simon}},\
	}\href {\doibase https://doi.org/10.1016/0001-8708(77)90108-6} {\bibfield
		{journal} {\bibinfo  {journal} {Advances in Mathematics}\ }\textbf {\bibinfo
			{volume} {23}},\ \bibinfo {pages} {22 } (\bibinfo {year} {1977})}\BibitemShut
	{NoStop}%
	\bibitem [{\citenamefont {Scott}(1952)}]{scott1952}%
	\BibitemOpen
	\bibfield  {author} {\bibinfo {author} {\bibfnamefont {J.}~\bibnamefont
			{Scott}},\ }\href {\doibase 10.1080/14786440808520234} {\bibfield  {journal}
		{\bibinfo  {journal} {Philosophical Magazine}\ }\textbf {\bibinfo {volume}
			{43}},\ \bibinfo {pages} {859} (\bibinfo {year} {1952})}\BibitemShut
	{NoStop}%
	\bibitem [{\citenamefont {Schwinger}(1980)}]{schwinger1980thomas}%
	\BibitemOpen
	\bibfield  {author} {\bibinfo {author} {\bibfnamefont {J.}~\bibnamefont
			{Schwinger}},\ }\href {\doibase 10.1103/PhysRevA.22.1827} {\bibfield
		{journal} {\bibinfo  {journal} {Phys. Rev. A}\ }\textbf {\bibinfo {volume}
			{22}},\ \bibinfo {pages} {1827} (\bibinfo {year} {1980})}\BibitemShut
	{NoStop}%
	\bibitem [{\citenamefont {Schwinger}(1981)}]{schwinger1981thomas}%
	\BibitemOpen
	\bibfield  {author} {\bibinfo {author} {\bibfnamefont {J.}~\bibnamefont
			{Schwinger}},\ }\href {\doibase 10.1103/PhysRevA.24.2353} {\bibfield
		{journal} {\bibinfo  {journal} {Phys. Rev. A}\ }\textbf {\bibinfo {volume}
			{24}},\ \bibinfo {pages} {2353} (\bibinfo {year} {1981})}\BibitemShut
	{NoStop}%
	\bibitem [{\citenamefont {Spruch}(1991)}]{spruch1991pedagogic}%
	\BibitemOpen
	\bibfield  {author} {\bibinfo {author} {\bibfnamefont {L.}~\bibnamefont
			{Spruch}},\ }\href {\doibase 10.1103/RevModPhys.63.151} {\bibfield  {journal}
		{\bibinfo  {journal} {Rev. Mod. Phys.}\ }\textbf {\bibinfo {volume} {63}},\
		\bibinfo {pages} {151} (\bibinfo {year} {1991})}\BibitemShut {NoStop}%
	\bibitem [{\citenamefont {Salasnich}(2007)}]{kirzhnitsND}%
	\BibitemOpen
	\bibfield  {author} {\bibinfo {author} {\bibfnamefont {L.}~\bibnamefont
			{Salasnich}},\ }\href {\doibase 10.1088/1751-8113/40/33/004} {\bibfield
		{journal} {\bibinfo  {journal} {Journal of Physics A: Mathematical and
				Theoretical}\ }\textbf {\bibinfo {volume} {40}},\ \bibinfo {pages} {9987}
		(\bibinfo {year} {2007})}\BibitemShut {NoStop}%
	\bibitem [{\citenamefont {Hodges}(1973)}]{hodges73}%
	\BibitemOpen
	\bibfield  {author} {\bibinfo {author} {\bibfnamefont {C.~H.}\ \bibnamefont
			{Hodges}},\ }\href@noop {} {\bibfield  {journal} {\bibinfo  {journal}
			{Canadian J. Phys}\ }\textbf {\bibinfo {volume} {51}},\ \bibinfo {pages}
		{1428} (\bibinfo {year} {1973})}\BibitemShut {NoStop}%
	\bibitem [{\citenamefont {Jones}\ and\ \citenamefont
		{Gunnarsson}(1989)}]{JandG}%
	\BibitemOpen
	\bibfield  {author} {\bibinfo {author} {\bibfnamefont {R.~O.}\ \bibnamefont
			{Jones}}\ and\ \bibinfo {author} {\bibfnamefont {O.}~\bibnamefont
			{Gunnarsson}},\ }\href {\doibase 10.1103/RevModPhys.61.689} {\bibfield
		{journal} {\bibinfo  {journal} {Rev. Mod. Phys.}\ }\textbf {\bibinfo {volume}
			{61}},\ \bibinfo {pages} {689} (\bibinfo {year} {1989})}\BibitemShut
	{NoStop}%
	\bibitem [{\citenamefont {Lindmaa}, \citenamefont {Mattsson},\ and\
		\citenamefont {Armiento}(2014)}]{lindmaa14}%
	\BibitemOpen
	\bibfield  {author} {\bibinfo {author} {\bibfnamefont {A.}~\bibnamefont
			{Lindmaa}}, \bibinfo {author} {\bibfnamefont {A.~E.}\ \bibnamefont
			{Mattsson}}, \ and\ \bibinfo {author} {\bibfnamefont {R.}~\bibnamefont
			{Armiento}},\ }\href {\doibase 10.1103/PhysRevB.90.075139} {\bibfield
		{journal} {\bibinfo  {journal} {Phys. Rev. B}\ }\textbf {\bibinfo {volume}
			{90}},\ \bibinfo {pages} {075139} (\bibinfo {year} {2014})}\BibitemShut
	{NoStop}%
	\bibitem [{\citenamefont {Mattsson}\ and\ \citenamefont
		{Kohn}(2001)}]{KohnMattsson}%
	\BibitemOpen
	\bibfield  {author} {\bibinfo {author} {\bibfnamefont {A.~E.}\ \bibnamefont
			{Mattsson}}\ and\ \bibinfo {author} {\bibfnamefont {W.}~\bibnamefont
			{Kohn}},\ }\href {\doibase 10.1063/1.1396649} {\bibfield  {journal} {\bibinfo
			{journal} {The Journal of Chemical Physics}\ }\textbf {\bibinfo {volume}
			{115}},\ \bibinfo {pages} {3441} (\bibinfo {year} {2001})},\ \Eprint
	{http://arxiv.org/abs/https://doi.org/10.1063/1.1396649}
	{https://doi.org/10.1063/1.1396649} \BibitemShut {NoStop}%
	\bibitem [{\citenamefont {Astakhov}, \citenamefont {Stash},\ and\ \citenamefont
		{Tsirelson}(2016)}]{Ts}%
	\BibitemOpen
	\bibfield  {author} {\bibinfo {author} {\bibfnamefont {A.~A.}\ \bibnamefont
			{Astakhov}}, \bibinfo {author} {\bibfnamefont {A.~I.}\ \bibnamefont {Stash}},
		\ and\ \bibinfo {author} {\bibfnamefont {V.~G.}\ \bibnamefont {Tsirelson}},\
	}\href {\doibase 10.1002/qua.24957} {\bibfield  {journal} {\bibinfo
			{journal} {International Journal of Quantum Chemistry}\ }\textbf {\bibinfo
			{volume} {116}},\ \bibinfo {pages} {237} (\bibinfo {year} {2016})},\ \Eprint
	{http://arxiv.org/abs/https://onlinelibrary.wiley.com/doi/pdf/10.1002/qua.24957}
	{https://onlinelibrary.wiley.com/doi/pdf/10.1002/qua.24957} \BibitemShut
	{NoStop}%
	\bibitem [{\citenamefont {Heilmann}\ and\ \citenamefont
		{Lieb}(1995)}]{heilmannlieb}%
	\BibitemOpen
	\bibfield  {author} {\bibinfo {author} {\bibfnamefont {O.~J.}\ \bibnamefont
			{Heilmann}}\ and\ \bibinfo {author} {\bibfnamefont {E.~H.}\ \bibnamefont
			{Lieb}},\ }\href {\doibase 10.1103/PhysRevA.52.3628} {\bibfield  {journal}
		{\bibinfo  {journal} {Phys. Rev. A}\ }\textbf {\bibinfo {volume} {52}},\
		\bibinfo {pages} {3628} (\bibinfo {year} {1995})}\BibitemShut {NoStop}%
	\bibitem [{\citenamefont {Fuchs}\ and\ \citenamefont
		{Scheffler}(1999)}]{FHI98PP}%
	\BibitemOpen
	\bibfield  {author} {\bibinfo {author} {\bibfnamefont {M.}~\bibnamefont
			{Fuchs}}\ and\ \bibinfo {author} {\bibfnamefont {M.}~\bibnamefont
			{Scheffler}},\ }\href@noop {} {\bibfield  {journal} {\bibinfo  {journal}
			{Computer Physics Communications}\ }\textbf {\bibinfo {volume} {119}},\
		\bibinfo {pages} {67} (\bibinfo {year} {1999})}\BibitemShut {NoStop}%
	\bibitem [{\citenamefont {Perdew}, \citenamefont {Burke},\ and\ \citenamefont
		{Wang}(1996)}]{numGGA}%
	\BibitemOpen
	\bibfield  {author} {\bibinfo {author} {\bibfnamefont {J.~P.}\ \bibnamefont
			{Perdew}}, \bibinfo {author} {\bibfnamefont {K.}~\bibnamefont {Burke}}, \
		and\ \bibinfo {author} {\bibfnamefont {Y.}~\bibnamefont {Wang}},\ }\href
	{\doibase 10.1103/PhysRevB.54.16533} {\bibfield  {journal} {\bibinfo
			{journal} {Phys. Rev. B}\ }\textbf {\bibinfo {volume} {54}},\ \bibinfo
		{pages} {16533} (\bibinfo {year} {1996})}\BibitemShut {NoStop}%
	\bibitem [{\citenamefont {Redd}()}]{thesis}%
	\BibitemOpen
	\bibfield  {author} {\bibinfo {author} {\bibfnamefont {J.~J.}\ \bibnamefont
			{Redd}},\ }\href@noop {} {\enquote {\bibinfo {title} {An analysis of atomic
				wave functions to improve density functional kinetic energy models},}\
	}\bibinfo {note} {Master's thesis, Ball State University (2015)}\BibitemShut
	{NoStop}%
	\bibitem [{\citenamefont {Pyykko}(2011)}]{pyykko}%
	\BibitemOpen
	\bibfield  {author} {\bibinfo {author} {\bibfnamefont {P.}~\bibnamefont
			{Pyykko}},\ }\href@noop {} {\bibfield  {journal} {\bibinfo  {journal} {Phys.
				Chem. Chem. Phys.}\ }\textbf {\bibinfo {volume} {13}},\ \bibinfo {pages}
		{161} (\bibinfo {year} {2011})}\BibitemShut {NoStop}%
	\bibitem [{\citenamefont {Engel}\ and\ \citenamefont {Dreizler}(1999)}]{opmks}%
	\BibitemOpen
	\bibfield  {author} {\bibinfo {author} {\bibfnamefont {E.}~\bibnamefont
			{Engel}}\ and\ \bibinfo {author} {\bibfnamefont {R.~M.}\ \bibnamefont
			{Dreizler}},\ }\href {\doibase
		10.1002/(SICI)1096-987X(19990115)20:1<31::AID-JCC6>3.0.CO;2-P} {\bibfield
		{journal} {\bibinfo  {journal} {Journal of Computational Chemistry}\ }\textbf
		{\bibinfo {volume} {20}},\ \bibinfo {pages} {31} (\bibinfo {year}
		{1999})}\BibitemShut {NoStop}%
	\bibitem [{\citenamefont {Acharya}\ \emph {et~al.}(1980)\citenamefont
		{Acharya}, \citenamefont {Bartolotti}, \citenamefont {Sears},\ and\
		\citenamefont {Parr}}]{acharya}%
	\BibitemOpen
	\bibfield  {author} {\bibinfo {author} {\bibfnamefont {P.~K.}\ \bibnamefont
			{Acharya}}, \bibinfo {author} {\bibfnamefont {L.~J.}\ \bibnamefont
			{Bartolotti}}, \bibinfo {author} {\bibfnamefont {S.~B.}\ \bibnamefont
			{Sears}}, \ and\ \bibinfo {author} {\bibfnamefont {R.~G.}\ \bibnamefont
			{Parr}},\ }\href@noop {} {\bibfield  {journal} {\bibinfo  {journal}
			{Proceedings of the National Academy of Sciences}\ }\textbf {\bibinfo
			{volume} {77}},\ \bibinfo {pages} {6978} (\bibinfo {year}
		{1980})}\BibitemShut {NoStop}%
	\bibitem [{\citenamefont {Constantin}, \citenamefont {Fabiano},\ and\
		\citenamefont {Sala}(2016)}]{constantin2016kinetic}%
	\BibitemOpen
	\bibfield  {author} {\bibinfo {author} {\bibfnamefont {L.~A.}\ \bibnamefont
			{Constantin}}, \bibinfo {author} {\bibfnamefont {E.}~\bibnamefont {Fabiano}},
		\ and\ \bibinfo {author} {\bibfnamefont {F.~D.}\ \bibnamefont {Sala}},\
	}\href@noop {} {\bibfield  {journal} {\bibinfo  {journal} {Computation}\
		}\textbf {\bibinfo {volume} {4}},\ \bibinfo {pages} {19} (\bibinfo {year}
		{2016})}\BibitemShut {NoStop}%
	\bibitem [{\citenamefont {Becke}\ and\ \citenamefont
		{Edgecombe}(1990)}]{BeckeEdgecombe}%
	\BibitemOpen
	\bibfield  {author} {\bibinfo {author} {\bibfnamefont {A.~D.}\ \bibnamefont
			{Becke}}\ and\ \bibinfo {author} {\bibfnamefont {K.~E.}\ \bibnamefont
			{Edgecombe}},\ }\href {\doibase http://dx.doi.org/10.1063/1.458517}
	{\bibfield  {journal} {\bibinfo  {journal} {J. Chem. Phys.}\ }\textbf
		{\bibinfo {volume} {92}},\ \bibinfo {pages} {5397} (\bibinfo {year}
		{1990})}\BibitemShut {NoStop}%
	\bibitem [{\citenamefont {Becke}(1998)}]{Becke98}%
	\BibitemOpen
	\bibfield  {author} {\bibinfo {author} {\bibfnamefont {A.~D.}\ \bibnamefont
			{Becke}},\ }\href@noop {} {\bibfield  {journal} {\bibinfo  {journal} {J.
				Chem. Phys.}\ }\textbf {\bibinfo {volume} {109}},\ \bibinfo {pages} {2092}
		(\bibinfo {year} {1998})}\BibitemShut {NoStop}%
	\bibitem [{\citenamefont {Sun}, \citenamefont {Xiao},\ and\ \citenamefont
		{Ruzsinszky}(2012)}]{SXR12}%
	\BibitemOpen
	\bibfield  {author} {\bibinfo {author} {\bibfnamefont {J.}~\bibnamefont
			{Sun}}, \bibinfo {author} {\bibfnamefont {B.}~\bibnamefont {Xiao}}, \ and\
		\bibinfo {author} {\bibfnamefont {A.}~\bibnamefont {Ruzsinszky}},\ }\href
	{\doibase 10.1063/1.4742312} {\bibfield  {journal} {\bibinfo  {journal} {The
				Journal of Chemical Physics}\ }\textbf {\bibinfo {volume} {137}},\ \bibinfo
		{pages} {051101} (\bibinfo {year} {2012})},\ \Eprint
	{http://arxiv.org/abs/https://doi.org/10.1063/1.4742312}
	{https://doi.org/10.1063/1.4742312} \BibitemShut {NoStop}%
	\bibitem [{\citenamefont {van Leeuwen}\ and\ \citenamefont
		{Baerends}(1994)}]{van1994exchange}%
	\BibitemOpen
	\bibfield  {author} {\bibinfo {author} {\bibfnamefont {R.}~\bibnamefont {van
				Leeuwen}}\ and\ \bibinfo {author} {\bibfnamefont {E.~J.}\ \bibnamefont
			{Baerends}},\ }\href {\doibase 10.1103/PhysRevA.49.2421} {\bibfield
		{journal} {\bibinfo  {journal} {Phys. Rev. A}\ }\textbf {\bibinfo {volume}
			{49}},\ \bibinfo {pages} {2421} (\bibinfo {year} {1994})}\BibitemShut
	{NoStop}%
	\bibitem [{\citenamefont {Perdew}\ and\ \citenamefont
		{Zunger}(1981)}]{PerdewZunger}%
	\BibitemOpen
	\bibfield  {author} {\bibinfo {author} {\bibfnamefont {J.~P.}\ \bibnamefont
			{Perdew}}\ and\ \bibinfo {author} {\bibfnamefont {A.}~\bibnamefont
			{Zunger}},\ }\href {\doibase 10.1103/PhysRevB.23.5048} {\bibfield  {journal}
		{\bibinfo  {journal} {Phys. Rev. B}\ }\textbf {\bibinfo {volume} {23}},\
		\bibinfo {pages} {5048} (\bibinfo {year} {1981})}\BibitemShut {NoStop}%
	\bibitem [{\citenamefont {Morrell}, \citenamefont {Parr},\ and\ \citenamefont
		{Levy}(1975)}]{mparrlevy}%
	\BibitemOpen
	\bibfield  {author} {\bibinfo {author} {\bibfnamefont {M.~M.}\ \bibnamefont
			{Morrell}}, \bibinfo {author} {\bibfnamefont {R.~G.}\ \bibnamefont {Parr}}, \
		and\ \bibinfo {author} {\bibfnamefont {M.}~\bibnamefont {Levy}},\ }\href
	{\doibase 10.1063/1.430509} {\bibfield  {journal} {\bibinfo  {journal} {The
				Journal of Chemical Physics}\ }\textbf {\bibinfo {volume} {62}},\ \bibinfo
		{pages} {549} (\bibinfo {year} {1975})},\ \Eprint
	{http://arxiv.org/abs/https://aip.scitation.org/doi/pdf/10.1063/1.430509}
	{https://aip.scitation.org/doi/pdf/10.1063/1.430509} \BibitemShut {NoStop}%
	\bibitem [{eva()}]{evanescent}%
	\BibitemOpen
	\href@noop {} {}\bibinfo {note} {This assumption is not true for orbitals
		derived from a scheme that involves the {F}ock operator, such as
		{H}artree-Fock~\cite{hms69} or many-body quasiparticle
		orbitals.~\cite{almbladhvonbarth} In this case all orbitals are coupled to
		each other and all will have an asymptotic piece that decays like the HOAO.
		But our analysis holds for orbitals derived from any method such as a GGA or
		metaGGA which could presumably be used in an orbital-free
		context.}\BibitemShut {Stop}%
	\bibitem [{\citenamefont {Sala}, \citenamefont {Fabiano},\ and\ \citenamefont
		{Constantin}(2015)}]{constantin2015atom}%
	\BibitemOpen
	\bibfield  {author} {\bibinfo {author} {\bibfnamefont {F.~D.}\ \bibnamefont
			{Sala}}, \bibinfo {author} {\bibfnamefont {E.}~\bibnamefont {Fabiano}}, \
		and\ \bibinfo {author} {\bibfnamefont {L.~A.}\ \bibnamefont {Constantin}},\
	}\href {\doibase 10.1103/PhysRevB.91.035126} {\bibfield  {journal} {\bibinfo
			{journal} {Physical Review B}\ }\textbf {\bibinfo {volume} {91}},\ \bibinfo
		{pages} {035126} (\bibinfo {year} {2015})}\BibitemShut {NoStop}%
	\bibitem [{\citenamefont {Giuliani}\ \emph {et~al.}(2019)\citenamefont
		{Giuliani}, \citenamefont {Matheson}, \citenamefont {Nazarewicz},
		\citenamefont {Olsen}, \citenamefont {Reinhard}, \citenamefont {Sadhukhan},
		\citenamefont {Schuetrumpf}, \citenamefont {Schunck},\ and\ \citenamefont
		{Schwerdtfeger}}]{Oganesson}%
	\BibitemOpen
	\bibfield  {author} {\bibinfo {author} {\bibfnamefont {S.~A.}\ \bibnamefont
			{Giuliani}}, \bibinfo {author} {\bibfnamefont {Z.}~\bibnamefont {Matheson}},
		\bibinfo {author} {\bibfnamefont {W.}~\bibnamefont {Nazarewicz}}, \bibinfo
		{author} {\bibfnamefont {E.}~\bibnamefont {Olsen}}, \bibinfo {author}
		{\bibfnamefont {P.-G.}\ \bibnamefont {Reinhard}}, \bibinfo {author}
		{\bibfnamefont {J.}~\bibnamefont {Sadhukhan}}, \bibinfo {author}
		{\bibfnamefont {B.}~\bibnamefont {Schuetrumpf}}, \bibinfo {author}
		{\bibfnamefont {N.}~\bibnamefont {Schunck}}, \ and\ \bibinfo {author}
		{\bibfnamefont {P.}~\bibnamefont {Schwerdtfeger}},\ }\href {\doibase
		10.1103/RevModPhys.91.011001} {\bibfield  {journal} {\bibinfo  {journal}
			{Rev. Mod. Phys.}\ }\textbf {\bibinfo {volume} {91}},\ \bibinfo {pages}
		{011001} (\bibinfo {year} {2019})}\BibitemShut {NoStop}%
	\bibitem [{\citenamefont {Jerabek}\ \emph {et~al.}(2018)\citenamefont
		{Jerabek}, \citenamefont {Schuetrumpf}, \citenamefont {Schwerdtfeger},\ and\
		\citenamefont {Nazarewicz}}]{Jerabek2018}%
	\BibitemOpen
	\bibfield  {author} {\bibinfo {author} {\bibfnamefont {P.}~\bibnamefont
			{Jerabek}}, \bibinfo {author} {\bibfnamefont {B.}~\bibnamefont
			{Schuetrumpf}}, \bibinfo {author} {\bibfnamefont {P.}~\bibnamefont
			{Schwerdtfeger}}, \ and\ \bibinfo {author} {\bibfnamefont {W.}~\bibnamefont
			{Nazarewicz}},\ }\href {\doibase 10.1103/PhysRevLett.120.053001} {\bibfield
		{journal} {\bibinfo  {journal} {Phys. Rev. Lett.}\ }\textbf {\bibinfo
			{volume} {120}},\ \bibinfo {pages} {053001} (\bibinfo {year}
		{2018})}\BibitemShut {NoStop}%
	\bibitem [{\citenamefont {Handy}, \citenamefont {Marron},\ and\ \citenamefont
		{Silverstone}(1969)}]{hms69}%
	\BibitemOpen
	\bibfield  {author} {\bibinfo {author} {\bibfnamefont {N.~C.}\ \bibnamefont
			{Handy}}, \bibinfo {author} {\bibfnamefont {M.~T.}\ \bibnamefont {Marron}}, \
		and\ \bibinfo {author} {\bibfnamefont {H.~J.}\ \bibnamefont {Silverstone}},\
	}\href {\doibase 10.1103/PhysRev.180.45} {\bibfield  {journal} {\bibinfo
			{journal} {Phys. Rev.}\ }\textbf {\bibinfo {volume} {180}},\ \bibinfo {pages}
		{45} (\bibinfo {year} {1969})}\BibitemShut {NoStop}%
	\bibitem [{\citenamefont {Almbladh}\ and\ \citenamefont {von
			Barth}(1985)}]{almbladhvonbarth}%
	\BibitemOpen
	\bibfield  {author} {\bibinfo {author} {\bibfnamefont {C.-O.}\ \bibnamefont
			{Almbladh}}\ and\ \bibinfo {author} {\bibfnamefont {U.}~\bibnamefont {von
				Barth}},\ }\href {\doibase 10.1103/PhysRevB.31.3231} {\bibfield  {journal}
		{\bibinfo  {journal} {Phys. Rev. B}\ }\textbf {\bibinfo {volume} {31}},\
		\bibinfo {pages} {3231} (\bibinfo {year} {1985})}\BibitemShut {NoStop}%
\end{thebibliography}
%

\end{document}